\documentclass[11pt,a4paper]{article}
\pdfoutput=1
\usepackage{jcappub}
\usepackage{rotating}
\usepackage{array}
\usepackage{amsmath}
\usepackage[normalem]{ulem}
\usepackage{slashed}
\usepackage{booktabs}
\usepackage[pdftex,table]{xcolor}
\usepackage{units}

\usepackage{multirow}

\arxivnumber{P3H-20-062, TTK-20-38}

\title{Resonant Sub-GeV Dirac Dark Matter}

\author{Elias Bernreuther,}
\author{Saniya Heeba,}
\author{and Felix Kahlhoefer}

\affiliation{Institute for Theoretical Particle Physics and Cosmology (TTK), RWTH Aachen University, \\ D-52056 Aachen, Germany}

\emailAdd{ebernreuther@physik.rwth-aachen.de}
\emailAdd{heeba@physik.rwth-aachen.de}
\emailAdd{kahlhoefer@physik.rwth-aachen.de}

\abstract{We study the phenomenology and detection prospects of a sub-GeV Dirac dark matter candidate with resonantly enhanced annihilations via a dark photon mediator. The model evades cosmological constraints on light thermal particles in the early universe while simultaneously being in reach of current and upcoming terrestrial experiments. We conduct a global analysis of the parameter space, considering bounds from accelerator and direct detection experiments, as well as those arising from Big Bang Nucleosynthesis, the Cosmic Microwave Background and dark matter self-interactions. We also extend our discussion to the case of a dark matter subcomponent. We find that large regions of parameter space remain viable even for the case of a moderate resonant enhancement, and demonstrate the complementarity of different experimental strategies for further exploring this scenario.}

\keywords{dark matter theory, dark matter experiments, cosmology of theories beyond the SM, particle physics - cosmology connection}

\begin{document}
\maketitle
\flushbottom

\section{Introduction}
\label{sec:introduction}

Recent years have witnessed an increasing interest in light (sub-GeV) dark matter (DM)~\cite{Knapen:2017xzo}. This has come on the heels of stringent direct detection constraints in the traditional (GeV--TeV) WIMP window along with significant improvements in experimental sensitivities, which continue to expand this window to smaller and smaller masses~\cite{Battaglieri:2017aum}. At the same time, accelerator experiments have made great advancements in the search for GeV-scale dark photons in both visible and invisible decay channels~\cite{Beacham:2019nyx,Fabbrichesi:2020wbt}. Since dark photons provide a well-motivated portal between the visible and dark sectors, light DM models with a dark photon mediator make for an attractive target that can be searched for in complimentary ways at direct/indirect detection and accelerator experiments. 

There are many ways in which sub-GeV DM coupled to a dark photon can reproduce the dark matter relic abundance measured by Planck~\cite{Aghanim:2018eyx}, including non-thermal production via the freeze-in mechanism~\cite{Hall:2009bx,Chu:2011be,Bernal:2017kxu,Heeba:2019jho}, annihilations into metastable states~\cite{Pospelov:2007mp} or number-changing processes~\cite{Hochberg:2014dra,Heeba:2018wtf} within the dark sector. Nevertheless, the conceptually simplest and most predictive scenario remains the conventional freeze-out mechanism, in which the DM relic abundance is determined by its annihilation cross section into Standard Model (SM) particles.

There are however strong constraints on sub-GeV DM particles in thermal contact with the SM. First of all, Big Bang Nucleosynthesis (BBN) places a stringent bound on the contribution of additional light degrees of freedom to the expansion rate of the Universe, which leads to a largely model-independent lower bound on the mass of thermal DM of about 10 MeV~\cite{Depta:2019lbe,Sabti:2019mhn}. But even heavier DM particles face strong constraints from the Cosmic Microwave Background (CMB), which is very sensitive to the injection of energy into the SM plasma due to the residual annihilations of DM at late times~\cite{Slatyer:2009yq,Slatyer:2015jla}. To evade these constraints it is necessary for models of thermal sub-GeV DM to suppress such late-time annihilations. This can be achieved for example if the annihilation of DM into SM particles is $p$-wave suppressed (as for complex scalar DM~\cite{Boehm:2003hm,Boehm:2020wbt}), if one annihilation partner is rapidly depleted (either due to a mass splitting as for inelastic DM~\cite{Izaguirre:2015zva,Duerr:2019dmv} or by an asymmetry in the dark sector~\cite{Kaplan:2009ag,Lin:2011gj}) or if the final state becomes kinematically inaccessible at low temperatures (as in forbidden DM~\cite{DAgnolo:2015ujb,Bernreuther:2019pfb}).

Comparably little attention has been devoted to the case that DM is a Dirac fermion, which at first sight has no mechanism to suppress CMB constraints. Nevertheless, a sufficiently strong velocity dependence can also be achieved in this case, if the annihilation cross section during freeze-out is resonantly enhanced. If this resonant enhancement is sufficiently large (i.e.\ if the relevant centre-of-mass energy is close to a narrow resonance), even tiny couplings can yield an acceptable relic abundance while completely evading experimental constraints~\cite{Feng:2017drg}. However, we will show that even a moderate enhancement, may be sufficient to satisfy cosmological constraints, while at the same time leading to potentially observable signals in existing and future laboratory experiments. 

Moreover, another way to evade the CMB constraint is to consider the case that sub-GeV DM only constitutes a fraction $R$ of the total DM abundance. In this case the annihilation rate is suppressed proportional to $R^2$, while event rates in direct detection experiments are suppressed proportional to $R$ and predictions for accelerator searches are unaffected~\cite{Chala:2015ama,Leane:2018kjk}. Hence, it becomes possible to satisfy constraints on the annihilation rate with relatively large (i.e.\ less finely tuned) couplings, which allow for observable rates in other types of experiments.

In order to comprehensively explore these effects, we perform a global analysis of Dirac DM coupled to a dark photon mediator. We conduct large-scale parameter scans, narrow down the viable regions of parameter space and identify the experiments suitable for further exploration of the model. We perform accurate relic density calculations following the formalism developed in ref.~\cite{Feng:2017drg} and use the state-of-the-art numerical tools \textsc{Hazma}~\cite{Coogan:2019qpu}, \textsc{HERWIG4DM}~\cite{Plehn:2019jeo} and \textsc{Darkcast}~\cite{Ilten:2018crw} to compute the relevant CMB and accelerator constraints. We also provide analytic expressions for our results wherever possible. 

We find that including resonant enhancement and the possibility of different DM sub-components leads to large viable regions of parameter space spanning many orders of magnitude in terms of the predicted event rates. Direct detection experiments are most sensitive to the case that DM is a sub-component with relatively little resonant enhancement. Accelerator searches prove to be highly complementary and probe the parameter space inaccessible to direct detection experiments.

The rest of this paper is structured as follows. In section~\ref{sec:set-up} we introduce the details of the model and establish the relevant notation. We then discuss the calculation of the thermally averaged annihilation cross section and the DM relic density. The relevant constraints from cosmology and terrestrial experiments as well as their technical implementation are discussed in sections~\ref{sec:cosmo} and~\ref{sec:terr_constraints}, respectively. Our results are presented in section~\ref{sec:results}, first for fixed parameter combinations and then using a global parameter scan. We conclude in section~\ref{sec:conclusions}. Additional details are provided in the appendices~\ref{sec:appendix_recasting} and~\ref{sec:appendix_eps}.

\section{General set-up}
\label{sec:set-up}

\subsection{Model details}
\label{sec:model}
We consider a Dirac fermion $\chi$ with mass $m_\chi$ coupled to a dark photon $A'$ with mass $m_{A'}$ that kinetically mixes with the SM photon. The $U(1)^\prime$ charge of $\chi$ guarantees its stability, making it an interesting DM candidate. We assume that the dark photon mass is generated via the Stueckelberg mechanism without the need for additional degrees of freedom~\cite{Stueckelberg:1900zz}.\footnote{We note in passing that the Stueckelberg mechanism can in principle give mass to any linear combination of the $U(1)^\prime$ and the $U(1)_Y$ gauge bosons, leading to an effective mass mixing between the two fields and ultimately millicharged DM~\cite{Feldman:2007wj}. In the present work we do not consider this possibility and instead assume that the Stueckelberg mechanism only gives mass to the $U(1)^\prime$ gauge boson. In this case the low-energy phenomenology is identical to the case where the dark photon mass is generated by a dark Higgs boson, except that there is no additional scalar particle in the spectrum.} After transforming the gauge fields to recover canonical kinetic terms, the Lagrangian of the model reads
\begin{align}
\mathcal{L}_\mathrm{DM} = -\frac{1}{4}F^\prime_{\mu\nu}F^{\prime\mu\nu} + \frac{1}{2}m_{A^\prime}^2A^{\prime2} -\kappa e A^\prime_\mu \sum_f q_f \bar{f} \gamma^\mu f + \bar{\chi}(i\slashed{\partial} - m_\chi)\chi - g_\chi A^\prime_\mu \bar{\chi} \gamma^\mu \chi\,,
\end{align}
where $\kappa$ denotes the kinetic mixing parameter and $g_\chi$ denotes the product of the $U(1)^\prime$ gauge coupling and the corresponding charge of $\chi$. The sum runs over all SM fermions with $q_f$ denoting the electromagnetic charge of $f$. We have omitted additional terms proportional to $\kappa \, m_{A^\prime}^2 / m_Z^2$ arising from mass mixing between the dark photon and the SM $Z$ boson, which have no phenomenological relevance.

The model we consider is hence fully characterised by the two masses $m_\chi$ and $m_{A^\prime}$ and the two couplings $g_\chi$ and $\kappa$. In the present work we will limit ourselves to the case that $m_\chi < m_{A^\prime} / 2$, such that dark photons can decay invisibly into a pair of DM particles. The decay width of the dark photon into a pair of leptons $\ell^+ \ell^-$ is given by
\begin{align}
\label{eq:widthSM}
\Gamma_\mathrm{\ell\ell} & = \frac{\kappa^2e^2m_{A^\prime}}{12\pi}\sqrt{1-\left(\frac{2m_\ell}{m_{A^\prime}}\right)^2}\left(1 + \frac{2m_\ell^2}{m_{A^\prime}^2}\right)\,,
\end{align}
such that the total visible decay width can be written as
\begin{align}
\Gamma_\mathrm{SM} &= R(m_{A^\prime}) \Gamma_{\mu\mu} + \sum_\ell \Gamma_{\ell\ell} \, ,
\label{eq:widthDM}
\end{align}
where 
$R(\sqrt{s})$ denotes the ratio of $\sigma(e^+ e^- \to \text{hadrons}) / \sigma(e^+ e^- \to \mu^+ \mu^-)$ via off-shell photons with centre-of-mass energy $\sqrt{s}$, which is resonantly enhanced if $\sqrt{s}$ is close to the mass of a QCD resonance with the appropriate quantum numbers~\cite{Ezhela:2003pp,Zyla:2020zbs}. The invisible decay width is given by
\begin{align}
\Gamma_\mathrm{DM} &=  \frac{g_\chi^2 m_{A^\prime}}{12\pi}\sqrt{1-\left(\frac{2m_\chi}{m_{A^\prime}}\right)^2}\left(1+\frac{2m_\chi^2}{m_{A^\prime}^2}\right)\,.
\end{align}
For future convenience we also define 
\begin{align}
 \Gamma_{A'} & = \Gamma_\mathrm{SM} + \Gamma_\mathrm{DM} \, ,\\ 
 \gamma_{A'} & = \frac{\Gamma_{A'}}{m_{A'}} \, , \\
 B_e & = \frac{\Gamma_{ee}}{\Gamma_\text{SM}} \; ,
\end{align}
which is the branching ratio into electrons of a dark photon with negligible invisible width.

In the following, we will focus on $10 \, \mathrm{MeV} \lesssim m_{A'} \lesssim 10 \, \mathrm{GeV}$. The lower bound is motivated by constraints from BBN (see section~\ref{sec:BBN}), while the upper bound is chosen such that the dark photon is kinematically accessible at $B$ factories, and the strong constraints from direct detection experiments for $m_\chi > 5 \, \mathrm{GeV}$ are evaded. In this mass range, the kinetic mixing parameter is constrained to $\kappa \lesssim 0.001$ irrespective of whether the dark photon decays dominantly visibly or invisibly. The coupling $g_\chi$, on the other hand, is only constrained by perturbativity: $g_\chi < \sqrt{4\pi}$. 

At the same time, for given $m_{A'}$ and $m_\chi$ one obtains lower bounds on $g_\chi$ and $\kappa$ from the requirement that DM particles annihilate away in the early Universe efficiently enough to be consistent with the observed DM relic abundance. The thermal DM relic density is set by processes of the kind $\chi\bar{\chi}\to A^\prime \to f\bar{f}$ which are efficient until the DM number density becomes Boltzmann suppressed and freezes out. For $m_\chi \ll m_{A'}$, the relic density requirement implies relatively large couplings, potentially in tension with CMB observations. However, if $m_\chi \approx m_{A'}/2$, annihilations are resonantly enhanced and an acceptable relic abundance can be achieved even for tiny couplings~\cite{Gondolo:1990dk, Feng:2017drg}. We parameterise the strength of the resonant enhancement by the dimensionless quantity
\begin{align}
\epsilon_R &= \frac{m_{A^\prime}^2 - 4 m_\chi^2}{4m_\chi^2} \,,
\end{align}
such that smaller values of $\epsilon_R$ correspond to larger enhancement.

\subsection{Relic density calculation}

Following ref.~\cite{Feng:2017drg}, we write the DM annihilation cross section as 
\begin{align}
\label{eq:sigmav_n}
\sigma v_\mathrm{lab} = F(\epsilon) \frac{m_{A^\prime} \Gamma_{A^\prime}}{(s-m_{A^\prime}^2 )^2 + m_{A^\prime}^2\Gamma_{A^\prime}^2}\,,
\end{align}
where $\epsilon = (s-4m_\chi^2)/4m_\chi^2$ is a dimensionless measure of kinetic energy of the collision, $v_\mathrm{lab} = 2\sqrt{\epsilon(1+\epsilon)}/(1+2\epsilon)$ is the relative velocity in the rest frame of one of the two particles
and
\begin{align}
F(\epsilon) = \frac{8 \pi  \alpha  \kappa ^2 g_{\chi }^2}{12 \,\pi \,\Gamma _{A^\prime}\, m_{A^\prime} \,m_{\chi }}\frac{(2 \epsilon +3) \left(m_e^2+2 (\epsilon +1) m_{\chi }^2\right) \sqrt{(\epsilon +1) m_{\chi }^2-m_e^2}}{(2 \epsilon +1) \sqrt{\epsilon +1} B_e\left(2 \sqrt{\epsilon +1} m_{\chi }\right)}\,,
\end{align}
where $\alpha = e^2/(4\pi)$ is the electromagnetic fine-structure constant. Note that in principle the width $\Gamma_{A^\prime}$ can depend on $\epsilon$ but this effect is small for cases of practical interest. Hence, we will approximate $\Gamma_{A^\prime}$ by the on-shell width evaluated at $m_{A^\prime}$ in the following.

The thermally averaged cross section is then given by
\begin{align}
\langle \sigma v \rangle = \frac{2 x}{K_2^2(x)}\int_0^{\infty} \,\sigma v \sqrt{\epsilon} (1+ 2\epsilon) K_1(2x\sqrt{1+\epsilon}) \,\mathrm{d}\epsilon \; .
\end{align}
For the calculation of thermal freeze-out it is usually a good approximation to take the non-relativistic limit, $\epsilon \ll 1$, which yields 
\begin{align}
\langle \sigma v \rangle = \frac{2 x^{3/2}}{\pi^{1/2}} \int_0^{\infty} \,\sigma v \sqrt{\epsilon} e^{-x\,\epsilon}\,\mathrm{d}\epsilon \; .
\end{align}
If we furthermore approximate $F(\epsilon) \approx F(0)$, the integration can be performed analytically to give
\begin{align}
\label{eq:sigmav}
\langle \sigma v \rangle_\mathrm{n.r.}  \approx \frac{ x^{3/2}\pi^{1/2}}{2m_\chi^2}\,F(0)\mathrm{Re}\left[z_R^{1/2} w(z_R^{1/2}x^{1/2})\right]\,,
\end{align}
where
\begin{align}
w(z) = \frac{2 i\,z}{\pi} \int_0^\infty \frac{e^{-t^2}}{z^2-t^2}\,\mathrm{d}t\,,
\end{align}
is the Fadeeva function and $z_R = \epsilon_R + i\,(1+\epsilon_R) \gamma_{A^\prime}$.

However, the non-relativistic limit is not a good approximation for $\epsilon_R \sim 1$ and $\gamma_{A'} \ll 1$, in which case a relevant contribution to the averaging integral comes from DM particles with relativistic velocities producing an on-shell dark photon ($\epsilon \approx \epsilon_R$). To capture this contribution, we can employ the narrow-width approximation:
\begin{equation}
\sigma v \approx \frac{\pi}{m_{A'}^2} F(\epsilon_R) \delta(\epsilon - \epsilon_R) 
\end{equation}
leading to
\begin{equation}
\langle \sigma v \rangle_\mathrm{res.} = \frac{\pi}{m_{A'}^2} F(\epsilon_R) \frac{2 x}{K_2^2(x)}\sqrt{\epsilon_R}\, (1+ 2\epsilon_R) K_1(2x\sqrt{1+\epsilon_R}) \; .
\end{equation}
At very low temperatures, the contribution from on-shell dark photon exchange is exponentially suppressed and becomes negligible compared to the contribution from off-shell dark photon exchange with $\epsilon < \epsilon_R$. The latter contribution can again be accurately captured by the non-relativistic limit discussed above.

\begin{figure}
\centering
\includegraphics[width=0.7\textwidth]{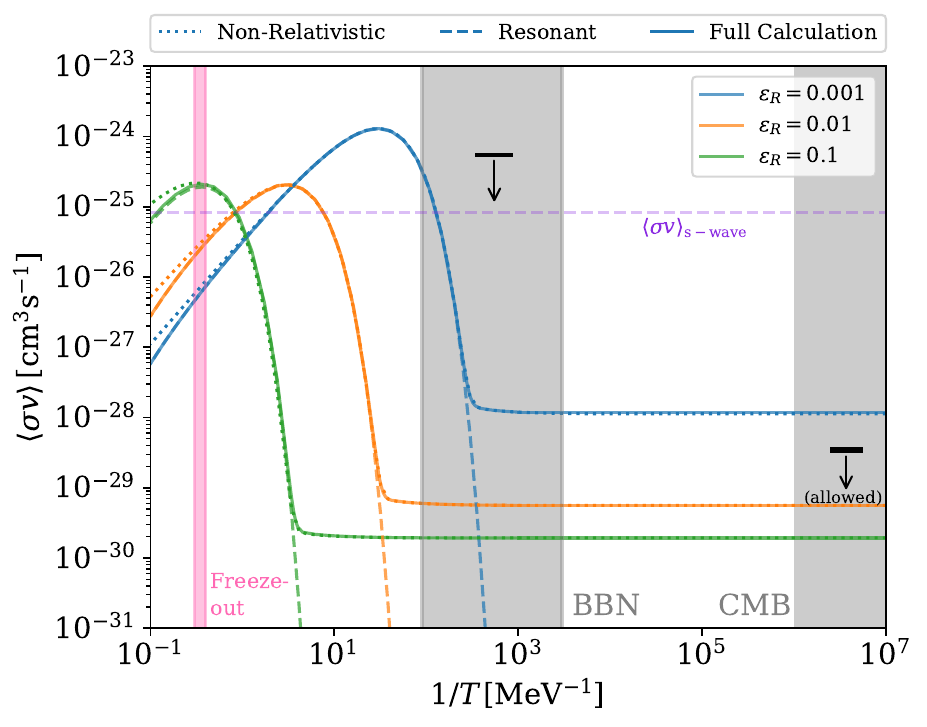}
\caption{\label{fig:sigmav}The thermally averaged cross section $\langle \sigma v \rangle$ as a function of inverse temperature for a DM mass $m_\chi = 50\,\mathrm{MeV}$, a DM coupling $g_\chi = 0.01$ and different values of the resonance parameter $\epsilon_R$. For each value of $\epsilon_R$ the kinetic mixing parameter $\kappa$ is fixed in such a way that the observed DM relic abundance is reproduced. Dotted (dashed) lines correspond to the non-relativistic approximation $\langle \sigma v\rangle_\text{n.r.}$ (resonant approximation $\langle \sigma v\rangle_\text{res.}$) as discussed in the text. We also indicate the approximate temperature ranges relevant for DM freeze-out, BBN and CMB. Horizontal bars indicate the annihilation cross sections required to satisfy the BBN constraint and the CMB constraint in the case of temperature-independent ($s$-wave) annihilations (indicated by the dotted purple line).}
\end{figure}

By comparing these approximate expressions to the full numerical result, we find that $\langle \sigma v \rangle_\mathrm{n.r.}$ gives a good approximation unless $\epsilon_R > 0.1$ and $\gamma_{A'} < 0.01$. In the latter case we use $\text{max}(\langle \sigma v \rangle_\mathrm{res.},\langle \sigma v \rangle_\text{n.r.})$ instead. 
Figure~\ref{fig:sigmav} shows the thermally averaged annihilation cross section as a function of inverse temperature together with the two analytic approximations introduced above for different values of $\epsilon_R$. For all curves we have fixed $m_\chi = 50\,\mathrm{MeV}$ and $g_\chi = 0.01$ and determined $\kappa$ by the requirement that the observed DM relic abundance is reproduced, which requires $\kappa = 3.6 \times 10^{-7}$, $8.0 \times 10^{-7}$ and $4.5 \times 10^{-6}$, for $\epsilon_R = 0.001$, $0.01$ and $0.1$, respectively (see below). As expected, we find that the non-relativistic approximation works very well for small $\epsilon_R$ as well as for very small temperatures, while the resonant approximation is more accurate for large $\epsilon_R$ and relatively large temperatures.

We also show in figure~\ref{fig:sigmav} the temperature ranges most relevant for DM freeze-out, BBN and the CMB together with the bounds on the DM annihilation cross section that one obtains from each of these epochs for $s$-wave annihilations into electrons (see section~\ref{sec:cosmo} for details). While the relic density and CMB constraints are clearly mutually exclusive for temperature-independent annihilations, they can be simultaneously satisfied in our set-up thanks to the resonant enhancement of the annihilation cross section during freeze-out.

The thermal relic abundance is given by 
\begin{align}
\label{eq:relic1}
\Omega h^2 &\approx 1.7 \times 10^{-10} \mathrm{GeV}^{-2} \left( \int_{x_f}^{x_0} \sqrt{g_\mathrm{eff}} \frac{\langle\sigma v\rangle}{x^2} \mathrm{d}x \right)^{-1}
\end{align}
with $g_\mathrm{eff}$ denoting the effective degrees of freedom present in the thermal plasma. Here the freeze-out temperature $x_f$ is defined by
\begin{align}
\frac{63 \sqrt{5} \,x_f^{-1/2} e^{-x_f} g}{32\pi^3} \frac{1}{g_\mathrm{eff}^{1/2}}m_\chi m_\mathrm{Pl} \langle \sigma v\rangle = 1\,,
\end{align}
where $g=2$ are the spin degrees of freedom of $\chi$ and $m_\mathrm{Pl} = 1.22 \times 10^{19}\,\mathrm{GeV}$ denotes the Planck mass.

We note that for $\epsilon_R \ll 1$ and $\gamma_{A'} \ll \epsilon_R$, it is possible to take the non-relativistic limit and make the narrow-width approximations simultaneously, which yields
\begin{align}
\label{eq:sigmav_res}
\langle \sigma v \rangle \approx \frac{2 \sqrt{\pi}x^{3/2}}{4m_\chi^2} \sqrt{\epsilon_R} e^{-x \epsilon_R} F(\epsilon_R)\,.
\end{align}
resulting in a relic abundance given by
\begin{align}
\label{eq:relic2}
\Omega h^2 =1.7 \times 10^{-10} \mathrm{GeV}^{-2} \frac{1}{g_\mathrm{eff}^{1/2}} \left( \frac{2\pi \,(1 + \epsilon_R) \,\mathrm{Erfc}(\sqrt{\epsilon_R x_f}) \,F(\epsilon_R)}{m_{A^\prime}^2}\right)^{-1}
\end{align}
We can use this result to make a few general observations about the model. It follows from eq.~(\ref{eq:relic2}) that
\begin{align}
\label{eq:relic3}
\Omega h^2 \propto \frac{1}{\mathrm{Erfc}(\sqrt{\epsilon_R x_f})\,F(\epsilon_{R})} \,\stackrel{\epsilon_R \ll1 }{\sim} \frac{\Gamma_{A'}}{\kappa^2\,g_\chi^2} \,.
\end{align} 
We conclude that for $g_\chi \gg \kappa$ ($\kappa \gg g_\chi$), the relic density becomes independent of  $g_\chi$ ($\kappa$). It is therefore possible to vary one of the couplings while keeping the relic density fixed, which increases viable parameter space of the model. 

We also note in passing that eq.~(\ref{eq:relic2}) implies that the relic density is essentially independent of $x_f$ for $\epsilon_R \ll 1$. The reason is that for $x_f < x < \epsilon_R^{-1}$ the thermally averaged cross section keeps growing and therefore the precise lower boundary of the integration is irrelevant (see figure~\ref{fig:sigmav}). On the other hand, for $\epsilon_R \sim 1$ our results are in agreement with the ones obtained by numerically integrating the cross section and solving the Boltzmann equation~\cite{Berlin:2018jbm,Duerr:2019dmv}.

As an example, for the parameter points represented in Fig.~\ref{fig:sigmav}, we obtain $\Omega h^2 = (0.121, 0.129, 0.122)$ when using eq.~(\ref{eq:relic1}) with numerically evaluated thermal averages. Using our approximate expressions of the thermally averaged cross sections, we obtain $\Omega h^2 = (0.126, 0.124, 0.128)$, while eq.~(\ref{eq:relic2}) yields $\Omega h^2 = (0.127, 0.124, 0.130)$ for $\epsilon_R = (0.001,0.01,0.1)$. The agreement between these values at the level of 5\% demonstrates the validity of our approximations.

To conclude this section, we emphasize that in the following we will be interested not only in parameter combinations that reproduce the observed relic density exactly, but also those that predict Dirac DM to be only a fraction of DM. We therefore define the fractional DM abundance
\begin{equation}
 R = \frac{\Omega h^2}{0.12} \; .
\end{equation}
We will use this factor to rescale some of the experimental bounds discussed below.

\section{Cosmological constraints}
\label{sec:cosmo}

\subsection{Cosmic Microwave Background}
\label{sec:CMB}

The strongest constraints on sub-GeV dark matter come from observations of the CMB. DM particles annihilating into SM particles during or after recombination inject energy into the photon-baryon plasma, which affects its temperature. The rate of energy injected per unit volume is characterised by the effective parameter 
\begin{align}
p_\mathrm{ann} = \frac{R^2}{2} f_\chi \frac{\langle \sigma v \rangle_\mathrm{CMB}}{m_\chi} \; ,
\end{align} 
where $f_\chi$ is the fraction of energy deposited in the plasma, which depends on the DM mass and the annihilation products, and  $\langle \sigma v \rangle_\mathrm{CMB}$ is the thermally averaged annihilation cross section at recombination. The factor $1/2$ accounts for the fact that DM is not self-conjugate in our model. The Planck upper bound on $p_\mathrm{ann}$ is $3.2 \times 10^{-28} \mathrm{cm}^3 \,\mathrm{s}^{-1}\,\mathrm{GeV}^{-1}$~\cite{Aghanim:2018eyx}. 

For an s-wave process, as is the case for our model, we can approximate $\langle \sigma v \rangle_\mathrm{CMB}$ by taking the limit $v \to 0$, which gives
\begin{align}
\label{eq:sigmav_CMB}
\langle \sigma v \rangle_\mathrm{CMB} = \frac{4\,\pi\alpha g_\chi^2\kappa^2}{m_{A^\prime}^3\,B_e\left(\tfrac{m_{A^\prime}}{\sqrt{1+\epsilon_R}}\right)}\,\frac{(1+\epsilon_R)^{3/2}\,(m_{A^\prime}^2+2\,(1+\epsilon_R)\,m_e^2)}{(1+\epsilon_R)^2\,\Gamma_{A^\prime}^2 + \epsilon_R^2\, m_{A^\prime}^2}\,\sqrt{\frac{m_{A^\prime}^2}{1+\epsilon_R}-4m_e^2} \; .
\end{align}
We note that for $1 \gg \epsilon_R \gg \gamma_{A'}$ this annihilation cross section is proportional to $\epsilon_R^{-2}$, while the relic density depends more weakly on $\epsilon_R$, see eq.~(\ref{eq:relic3}). As a result, CMB constraints typically get stronger for smaller $\epsilon_R$ (see also figure~\ref{fig:sigmav}).

To calculate $f_\chi$ we first of all need the energy injection spectra of positrons and photons from DM annihilations, $\mathrm{d}N_{e^+} / \mathrm{d} E$ and $\mathrm{d}N_\gamma / \mathrm{d} E$. These spectra can be calculated by weighting the energy injection spectra for individual annihilation channels with the appropriate branching ratios:
\begin{equation}
 \frac{\mathrm{d}N_{e^+,\gamma}}{\mathrm{d}E} = \frac{1}{\Gamma_\text{SM}} \left(\sum_\ell \Gamma^\ast_{\ell\ell} \left.\frac{\mathrm{d}N_{e^+,\gamma}}{\mathrm{d}E}\right\rvert_{\chi\bar{\chi}\to \ell^+ \ell^-} + R(2 m_\chi) \Gamma^\ast_{\mu\mu} \left.\frac{\mathrm{d}N_{e^+,\gamma}}{\mathrm{d}E}\right\rvert_{\chi\bar{\chi}\to q \bar{q}}  \right)\; ,
\end{equation}
where $\Gamma^\ast_{\ell\ell}$ denotes the partial decay width into $\ell^+ \ell^-$ of an off-shell dark photon with invariant mass $m_{A'^\ast} = 2 m_\chi$. 

While the injection spectra for electrons and muons in the final states can be calculated analytically, obtaining the injection spectra for taus and quarks is more involved. The well-known \textsc{PPPC} spectra~\cite{Cirelli:2010xx} are available only for $m_\chi > 5 \, \mathrm{GeV}$. Only very recently have the spectra for smaller masses been calculated. Ref.~\cite{Coogan:2019qpu} provides spectra for $m_\chi < 0.25 \, \mathrm{GeV}$, which are available via the public tool \textsc{Hazma}, while ref.~\cite{Plehn:2019jeo} provides spectra for the intermediate range $0.25 \, \mathrm{GeV} < m_\chi < 5 \, \mathrm{GeV}$, which will be made publicly available in future versions of \textsc{HERWIG}.\footnote{We thank Peter Reimitz for providing us the relevant injection spectra ahead of the public release.}

The energy injection spectra can then be convoluted with energy-dependent efficiency factors following ref.~\cite{Slatyer:2015jla}:
\begin{equation}
 f_\chi(m_\chi) = \frac{1}{2 m_\chi} \int_0^{m_\chi} \mathrm{d}E \, E \, \left[ 2 f_\text{eff}^{e^+}(E) \frac{\mathrm{d}N_{e^+}}{\mathrm{d}E} + f_\text{eff}^{\gamma}(E) \frac{\mathrm{d}N_{\gamma}}{\mathrm{d}E} \right] \; .
\end{equation}
Here we have made use of the fact that at the temperatures relevant for CMB constraints the resonant enhancement of the annihilation cross section is inefficient and the energy injection rate therefore has the same red-shift dependence as for the case of $s$-wave annihilations.

\begin{figure}
\centering
\includegraphics[width=0.6\textwidth]{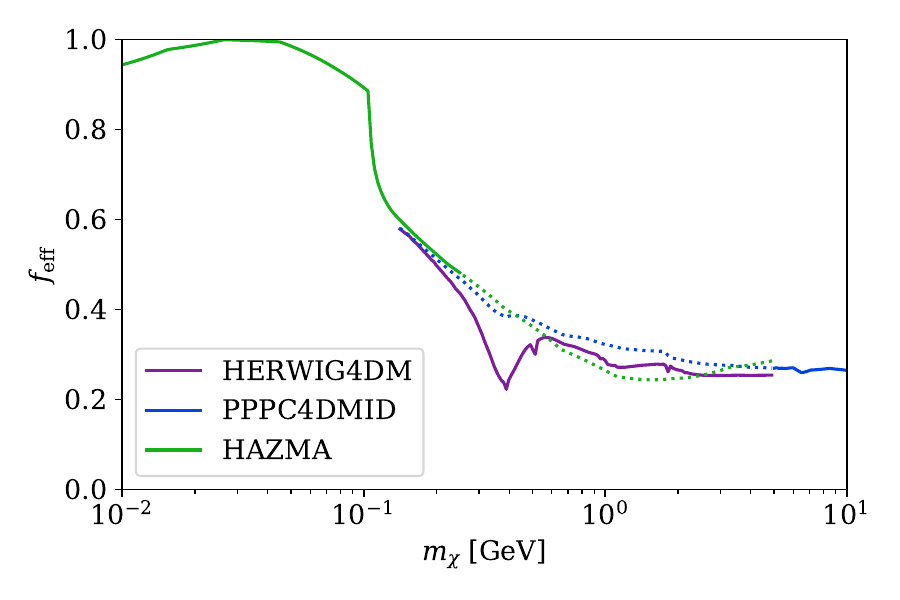}
\caption{\label{fig:feff} Efficiency function relevant for the calculation of CMB constraints as a function of the DM mass, calculated using the energy injection spectra from three different numerical tools. Dotted lines represent extrapolations beyond the range of validity of each code.}
\end{figure} 

Figure~\ref{fig:feff} shows the results of this procedure. We find that \textsc{Hazma} and \textsc{HERWIG4DM} give similar results for $m_\chi \approx 200 \, \mathrm{MeV}$, while \textsc{PPPC4DMID} and \textsc{HERWIG4DM} agree for $m_\chi > 2 \, \mathrm{GeV}$. However, the extrapolation of either \textsc{Hazma} or \textsc{PPPC4DMID} into the intermediate mass range (indicated by dotted lines) gives inaccurate results, as the details of the various QCD resonances are not captured. In the following, we will therefore use the results from $\textsc{Hazma}$ for $m_\chi < 200 \, \mathrm{MeV}$, from $\textsc{PPPC4DMID}$ for $m_\chi > 5 \, \mathrm{GeV}$ and from \textsc{HERWIG4DM} in the intermediate range. With this prescription we can convert the bound on $p_\mathrm{ann}$ into an upper bound on $R^2 \langle \sigma v \rangle_\mathrm{CMB}$ as a function of the DM mass.

\subsection{Big Bang Nucleosynthesis}
\label{sec:BBN}

The formation of light elements at temperatures below $1\,\mathrm{MeV}$ is sensitive to sub-GeV DM in two different ways. First, if the DM particles contribute significantly to the Hubble expansion rate, they will change the temperature of neutron decoupling and hence the helium abundance. This consideration places a lower bound on the mass of \emph{any} new particles that are in contact with the SM thermal bath (as well as an upper bound on the abundance of particles that are not in thermal contact with the SM). For the case of a Dirac fermion, this bound corresponds to approximately $m_\chi \gtrsim 10\,\mathrm{MeV}$~\cite{Depta:2019lbe,Sabti:2019mhn}.

The second effect is that DM annihilations at temperatures below $10\,\mathrm{keV}$ can lead to photodisintegration of light nuclei, in particular deuterium. For velocity-independent annihilations, this effect imposes the bound $\sigma v < 5.2 \times 10^{-25}\,\mathrm{cm^3/s}$~\cite{Depta:2019lbe}, i.e.\ it does not typically constrain cross sections compatible with the observed relic abundance. In our case, however, the cross section is substantially enhanced at low temperatures (see figure~\ref{fig:sigmav}).
In the following, we will therefore restrict ourselves to $\epsilon_R > 0.001$, which~-- together with the bound on $m_\chi$ discussed above~-- ensures that $\langle \sigma v \rangle$ is sufficiently suppressed at temperatures below $10\,\mathrm{keV}$ to evade BBN constraints. A detailed study of photodisintegration constraints will be left for future work.

\subsection{Dark matter self-interactions}
\label{sec:SIDM}

In the model that we consider, there are two types of DM self-interactions: particle-particle scattering via dark photon exchange in the $t$- or $u$-channel and particle-antiparticle scattering via dark photon exchange in the $s$- or $t$-channel. Since we are interested in the case where $m_{A'} \approx 2 m_\chi$, the $s$-channel exchange receives a resonant enhancement and therefore gives the dominant contribution to DM self-interactions. 
The quantity relevant for observational constraints is then the averaged momentum transfer cross section
\begin{equation}
 \overline{\sigma_T} = \frac{1}{2} \int_{-1}^1 (1 - |\cos \theta|) \frac{\mathrm{d}\sigma(\chi \bar{\chi}\to\chi\bar{\chi})}{\mathrm{d}\cos \theta} \mathrm{d}\cos\theta \; ,
\end{equation}
where the factor $\tfrac{1}{2}$ reflects the fact that only half of the collisions involve a particle-antiparticle pair.

The leading constraint on $\overline{\sigma_T}$ stems from the Bullet Cluster~\cite{Clowe:2006eq}, which probes DM velocities of the order of $\sigma \approx 1000 \, \mathrm{km/s} \approx 3 \times 10^{-3}c$. As long as $\sigma^2 \ll \epsilon_R$ we can calculate the self-interaction cross section in the limit $v \to 0$.
Including only the $s$-channel contribution, one obtains
\begin{equation}
 \overline{\sigma_T} = \frac{3 g_\chi^4 }{64 \pi [ 4 m_\chi^2 \epsilon_R^2 + (1 + \epsilon_R) \Gamma^2]} \; .
\end{equation}
The scattering rate of a particle in the Bullet Cluster is then given by
\begin{equation}
 \Gamma_\text{SIDM} = \overline{\sigma_T} \frac{R \rho_\text{BC}}{m_\chi} \; ,
\end{equation}
where $\rho_\text{BC}$ denotes the DM density in the Bullet Cluster and $R$ again denotes the fractional abundance of self-interacting DM particles.

To obtain the relative number of self-interacting DM particles lost during the cluster collision, we need to integrate the scattering rate across the entire Bullet Cluster. This leads to~\cite{Kahlhoefer:2013dca}
\begin{equation}
 \frac{\Delta N}{N} = 1 - e^{- R \, \overline{\sigma_T} \, \Sigma_\text{BC} / m_\chi} \; ,
\end{equation}
where $\Sigma_\text{BC} = \int \rho_\text{BC} \mathrm{d}{x} \approx 0.3 \, \mathrm{g / cm^2}$ denotes the surface density of the Bullet Cluster. Observations now require that the total amount of mass lost during the cluster collision is not greater than about 30\%~\cite{Markevitch:2003at}:
\begin{equation}
 R \frac{\Delta N}{N} < 0.3 \; ,
\end{equation}
where the additional factor of $R$ reflects the contribution of self-interacting DM to the total DM density. Note that for small self-interaction cross sections the effect is proportional to $R^2$, while for large self-interaction cross sections the total amount of mass lost can never exceed $R$. Hence, there is no constraint from the Bullet Cluster on a sub-dominant component of self-interacting DM with $R < 0.3$.

\section{Terrestrial constraints}
\label{sec:terr_constraints}

\subsection{Direct detection constraints}
\label{sec:DD}

Direct detection experiments provide some of the strongest bounds on DM--SM interactions. For the case of light DM, the relevant bounds are those which constrain the DM-electron scattering cross section given by~\cite{Essig:2011nj}
\begin{align}
\sigma_e = \frac{4 \mu_{\chi, e}^2 \,\alpha \kappa^2 g_\chi^2}{(m_{A^\prime}^2 + \alpha^2 m_e^2)^2}\,, 
\end{align}
where $\mu_{\chi,e} \equiv m_\chi m_e / (m_\chi + m_e) \approx m_e$ is the reduced mass of the DM-electron system.

For DM masses above about $100 \, \mathrm{MeV}$ there are also relevant constraints from experiments searching for DM-nucleon scattering. Bounds on the DM-nucleon cross section can be converted into those on $\sigma_e$ by a simple rescaling:
\begin{align}
\sigma_e = \frac{A^2}{Z^2} \frac{\mu_{\chi, e}^2}{\mu_{\chi, N}^2}\sigma_N\;,
\end{align}
where $\mu_{\chi, N}$ is the reduced mass of the DM-nucleon system, and $A$ and $Z$ denote respectively the mass and charge number of the target nucleus. This additional rescaling factor reflects the fact that bounds on spin-independent DM-nucleon scattering conventionally assume the cross section to be proportional to $A^2$, while for dark photon exchange it is proportional to $Z^2$.
Finally, bounds on the DM-nucleon scattering cross section can also be obtained by searching for ionisation electrons produced via the Migdal effect~\cite{Ibe:2017yqa,Dolan:2017xbu}. Again, these bounds can be rescaled to yield bounds on $\sigma_e$.

The leading bounds come from XENON1T \cite{Aprile_2019} and SENSEI \cite{Barak:2020fql} for DM-electron scattering and from CRESST-III \cite{Abdelhameed:2019hmk}, CDMSLite \cite{Agnese:2018gze} and a recent search for the Migdal effect in XENON1T~\cite{Aprile:2019jmx} for DM-nucleus scattering. To estimate the potential reach of near-future experiments, we will also consider sensitivity projections from SENSEI, DAMIC \cite{Aguilar-Arevalo:2019wdi}, SuperCDMS \cite{Amaral:2020ryn}, NEWS-G \cite{Arnaud:2017bjh} and CRESST-III~\cite{Reindl}, which have been obtained via the Dark Matter Limit Plotter~\cite{DMLP}.
Since the expected event rate at direct detection experiments is proportional to the local DM density, we interpret the exclusion limits in terms of $R\,\sigma_e$ to account for the case of a DM subcomponent.

\subsection{Accelerator constraints}
\label{sec:accelerator}

Dark photons with masses in the sub-GeV range can be produced in electron or proton beam dumps and other fixed-target experiments. Dark photon production modes in these experiments include bremsstrahlung off protons or electrons~\cite{Konaka:1986cb, Riordan:1987aw, Bjorken:1988as, Bross:1989mp, Davier:1989wz, Abrahamyan:2011gv, Merkel:2011ze, Merkel:2014avp, Banerjee:2016tad, Banerjee:2017hhz, Banerjee:2018vgk, Adrian:2018scb,  Banerjee:2019hmi, NA64:2019imj} as well as the decay of SM mesons with dark photons as decay products~\cite{Bergsma:1985qz, Bernardi:1985ny, Astier:2001ck,
Gninenko:2011uv,Gninenko:2012eq, Batley:2015lha,Tsai:2019mtm, CortinaGil:2019nuo}.\footnote{A third production mechanism based on the annihilation of positrons produced in the electromagnetic shower was recently discussed in Ref.~\cite{Marsicano:2018glj}.} The relevant final states are typically lepton pairs for visibly decaying dark photons and missing energy for invisible dark photons.

Moreover, MeV- to GeV-scale dark photons can be produced at $e^+ e^-$ colliders~\cite{Fox:2011fx, Babusci:2012cr, Babusci:2014sta, Anastasi:2015qla, Anastasi:2016ktq, Anastasi:2018azp, Lees:2014xha, Lees:2017lec,  Adachi:2019otg}. Searches for visible dark photons there are again typically sensitive to the decay into leptons, while invisible dark photons can give rise to missing energy in single-photon searches. Finally, visibly decaying GeV-scale dark photons are constrained by a number of searches at the LHC~\cite{Aaij:2017rft, Aaij:2019bvg, CMS:2016tgd, CMS:2018rdr, CMS:2019kiy}.

To recast fixed-target and collider constraints for our model we use the public code \textsc{Darkcast}, which implements the branching ratio calculation and recasting procedure from ref.~\cite{Ilten:2018crw}. We further modify \textsc{Darkcast} to account for the invisible dark photon decays present in our model. Details on the recasting are given in appendix~\ref{sec:appendix_recasting}.

Both at fixed-target experiments and at colliders the number of signal events $N$ is proportional to the product of the dark photon production cross section $\sigma_{A'}$, the branching ratio $\mathrm{BR}_{A'\to\mathcal{F}} \equiv \Gamma_\mathcal{F} / \Gamma_{A'}$  into the relevant final state and the experimental efficiency $\epsilon$:
\begin{align}
\label{eq:csbrepsilon}
N \propto \sigma_{A'} \; \mathrm{BR}_{A'\to\mathcal{F}} \; \epsilon \; .
\end{align}
The production cross section depends on the coupling of the dark photon to SM particles and is hence proportional to $\kappa^2$, while the efficiency $\epsilon$ of searches for long-lived dark photons depends sensitively on the lifetime $\tau_{A'} = \Gamma_{A'}^{-1}$ of the dark photon and its typical boost, which is determined by kinematics and depends on the dark photon mass $m_{A'}$.

We emphasize that the dark sector parameters $g_\chi$ and $\epsilon_R$ do not enter any of the terms in eq.~\eqref{eq:csbrepsilon} independently but only through the partial width $\Gamma_\mathrm{DM}$. In other words, different combinations of $g_\chi$ and $\epsilon_R$ that correspond to the same value of $\Gamma_\text{DM}$ yield the same experimental constraints on $\kappa$ and $m_{A'}$. Hence, it is useful to define the reduced partial width
\begin{align}
\gamma_\mathrm{inv} \equiv \frac{\Gamma_{\mathrm{DM}}}{m_A'} = \frac{g_\chi^2}{12\pi}\left(1-\frac{1}{1+\epsilon_R}\right)^{1/2}\left(1+\frac{1}{2(1+\epsilon_R)}\right) \; 
\end{align}
and use this parameter instead of $g_\chi$ as one of the four free parameters of our model.
For $\epsilon_R \ll 1$ the reduced partial width scales approximately like
\begin{align}
\gamma_\mathrm{inv}  \propto g_\chi^2 \; \sqrt{\epsilon_R} \; .
\end{align}

\section{Results}
\label{sec:results}

Having discussed all relevant observations and experiments that constrain the model under consideration, we are now in the position to determine the viable regions of parameter space. We will first do so by considering fixed values for two of the four model parameters and then present the results obtained when scanning over some or all model parameters simultaneously.

\subsection{Fixed parameter combinations}

\begin{figure}[t]
	\centering
	\includegraphics[height=6.5cm,clip,trim=15 5 90 5]{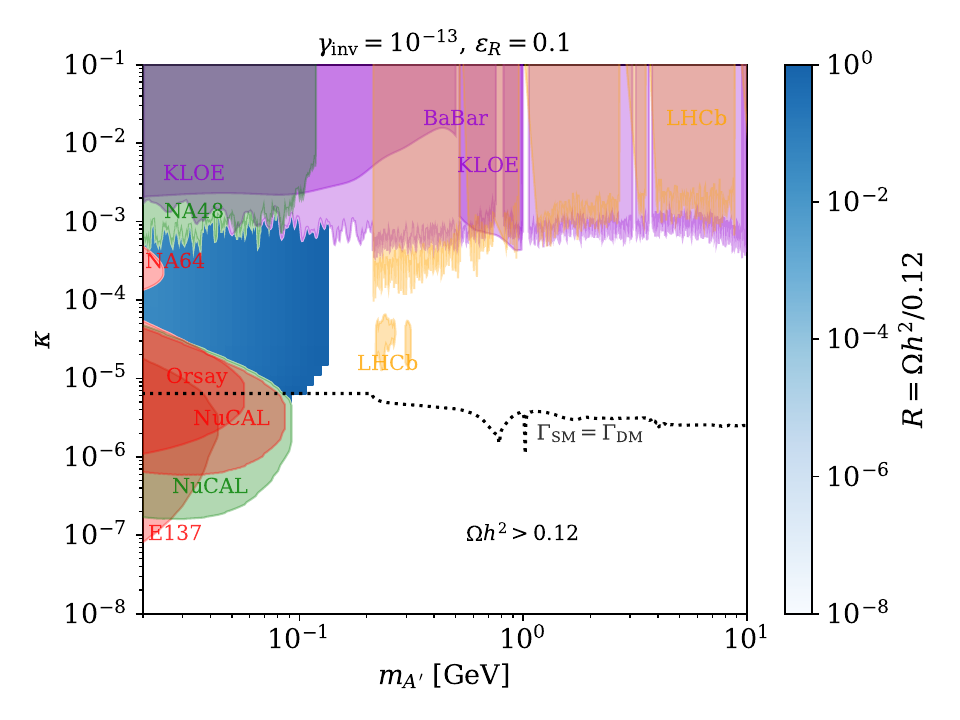}
	\includegraphics[height=6.5cm,clip,trim=15 5 15 5]{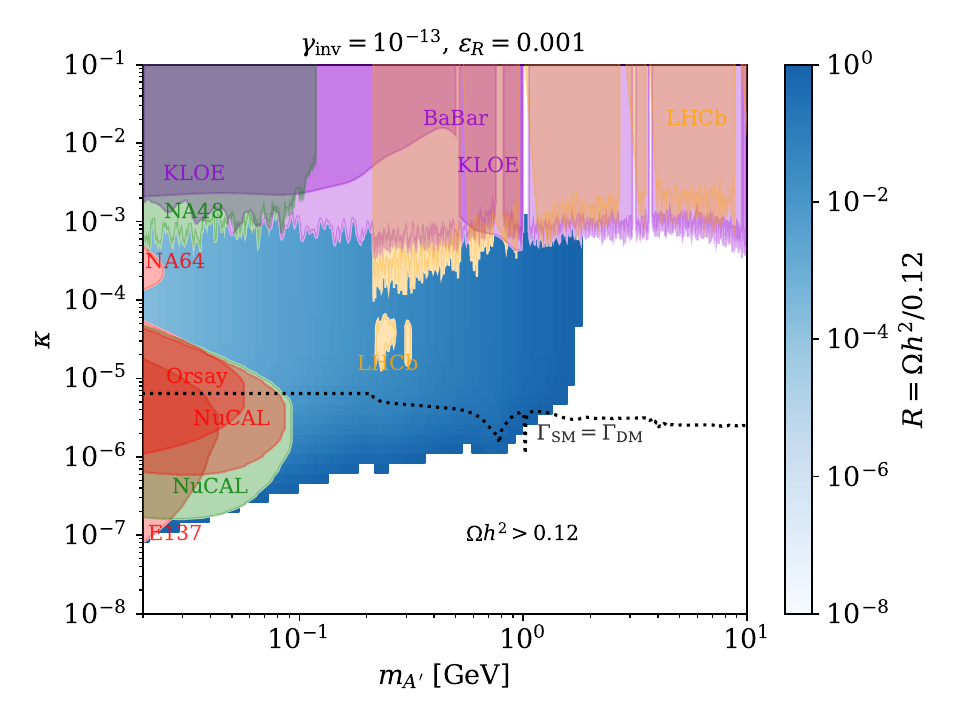}
	\includegraphics[height=6.5cm,clip,trim=15 5 90 5]{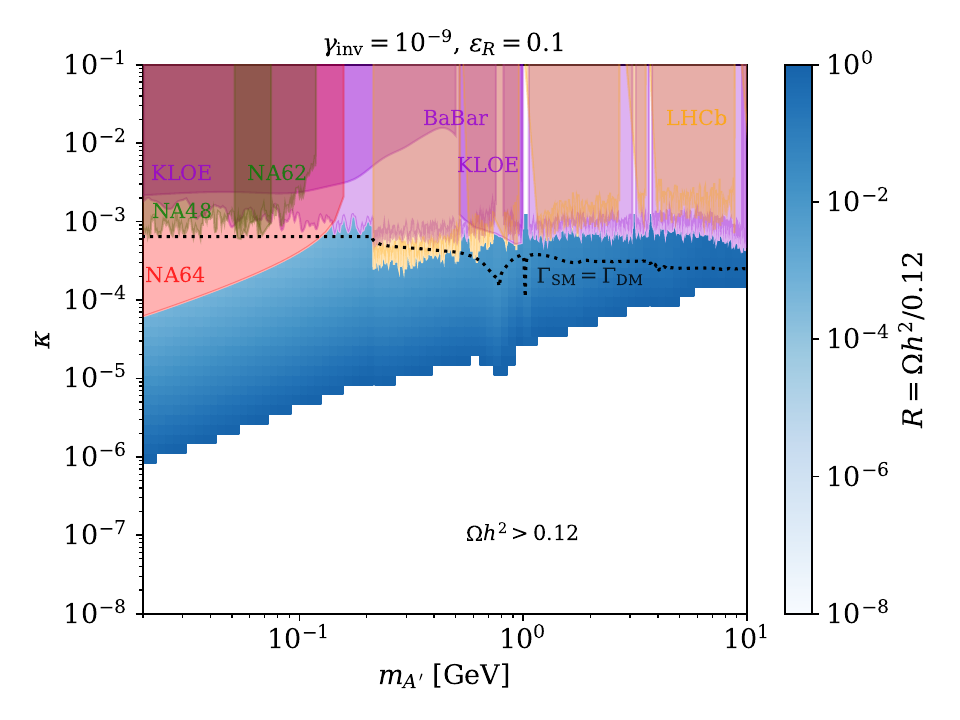}
	\includegraphics[height=6.5cm,clip,trim=15 5 15 5]{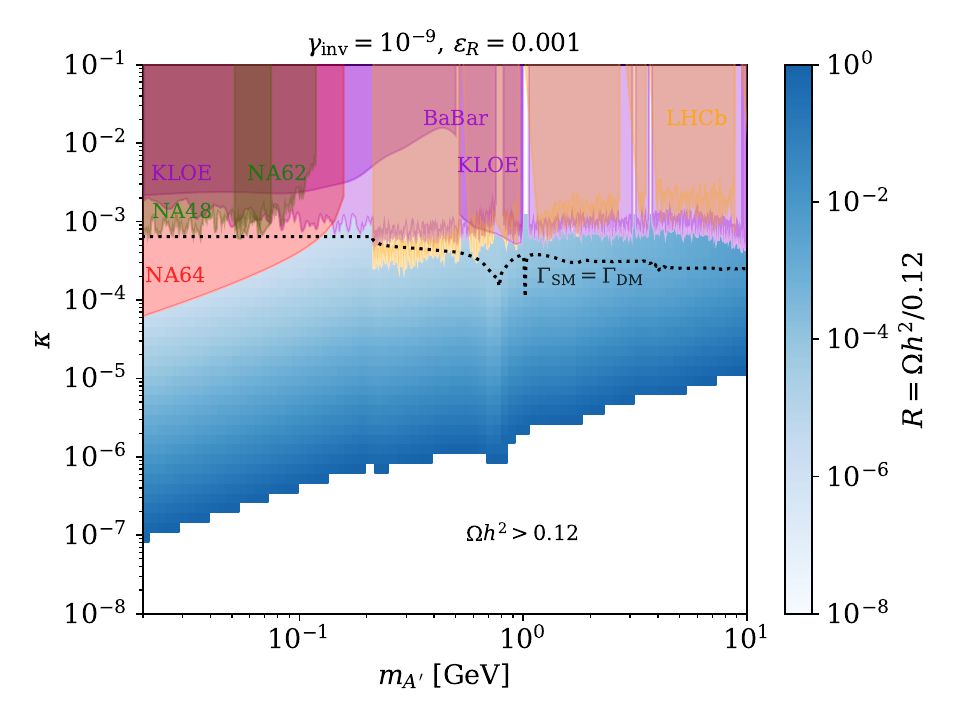}
	\includegraphics[height=6.5cm,clip,trim=15 5 90 5]{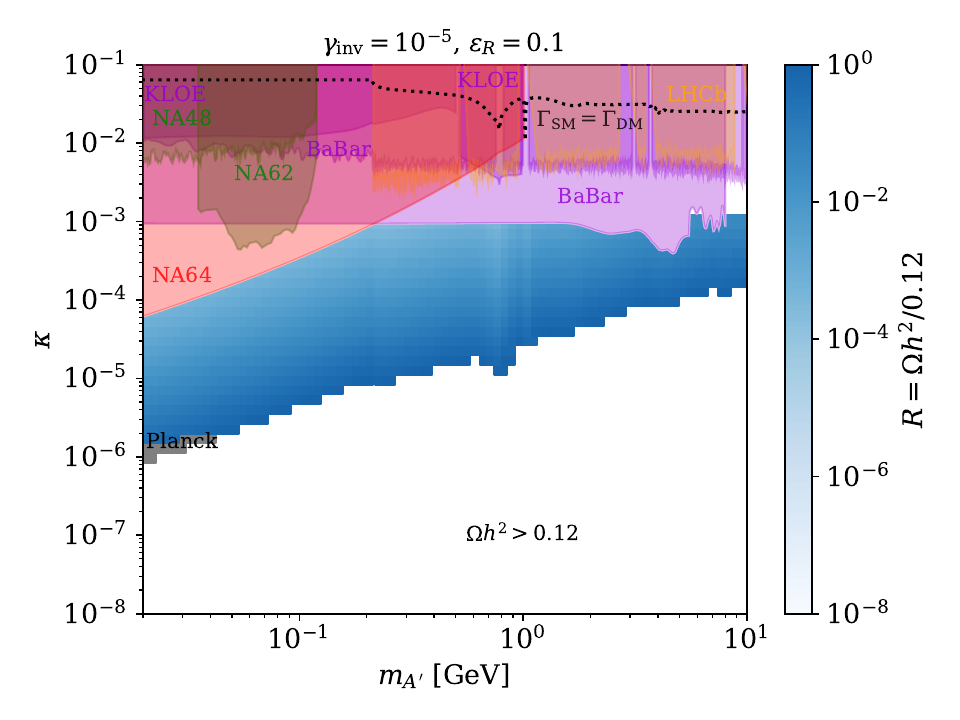}
	\includegraphics[height=6.5cm,clip,trim=15 5 15 5]{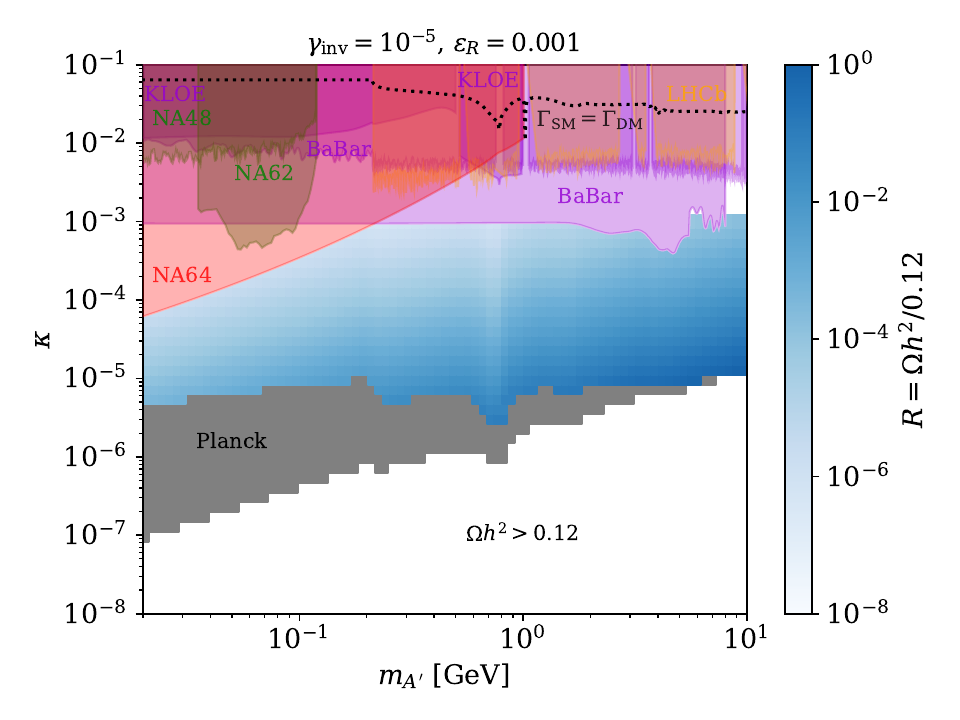}
	\caption{Experimental exclusion contours from fixed target and collider experiments for fixed $\gamma_\mathrm{inv}$ and $\epsilon_R$ on top of parameter points yielding $\Omega h^2\leq 0.12$. For each limit the dark photon production mechanism is indicated by the colour of the shading (red: bremsstrahlung, green: meson decay, violet: $e^+e^-$ colliders, orange: LHC). Grey shading indicates the parameter regions excluded by Planck.\label{fig:exp_excl_fixedeps}}
\end{figure}

In figure~\ref{fig:exp_excl_fixedeps}, we show the experimental exclusion contours calculated with \textsc{Darkcast} in the $m_{A'}$--$\kappa$ plane for different choices of $\gamma_\text{inv}$ and $\epsilon_R$. For each point we also calculate the relic density following section~\ref{sec:model}. For parameter combinations that yield $\Omega h^2 \leq 0.12$, we indicate the predicted relic density by blue shading, with dark blue representing the case that the DM particle constitutes most or all of DM and light blue points representing the case of a subdominant component. Parameter combinations that predict an unacceptably large relic abundance are excluded and shown without shading.

The different rows in figure~\ref{fig:exp_excl_fixedeps} correspond to different values of $\gamma_\text{inv}$. For each value of $\gamma_\mathrm{inv}$ we show in the two columns the predicted relic density for $\epsilon_{R}=0.1$ and \mbox{$\epsilon_{R}=0.001$}, corresponding to slight resonant enhancement and strong resonant enhancement of $\chi\bar{\chi}$ annihilation, respectively. Note that, as discussed above, for fixed $\gamma_\text{inv}$ the constraints from accelerator experiments are independent of $\epsilon_R$.

We first consider $\gamma_\mathrm{inv} = 10^{-13}$ (top row of figure~\ref{fig:exp_excl_fixedeps}), which corresponds to coupling values $g_\chi$ between $10^{-6}$ and $10^{-5}$ for the $\epsilon_R$ range considered here. Hence, the dark photon $A'$ decays predominantly into visible SM particles in the majority of the shown parameter space. Therefore, the strongest constraints for light dark photons with mass $m_{A'} \lesssim 200$~MeV and SM coupling $\kappa \lesssim 0.001$ come from fixed-target experiments searching for visible decays. For larger $\kappa$ almost the entire mass range between $10$~MeV and $10$~GeV is ruled out by a number of searches for visible final states at $e^+e^-$ colliders and at the LHC. For $\epsilon_R=0.1$ we only find viable parameter space for our model at dark photon masses below $200$~MeV. For $\epsilon_R=0.001$ the viable parameter space expands to larger dark photon masses and smaller $\kappa$ as the resonant enhancement of the DM annihilation cross section increases. As shown in eq.~\eqref{eq:relic3}, we have $\Omega h^2 \sim \Gamma_{A'}/(\kappa^2g_\chi^2)$ for strong resonant enhancement, i.e.\ for $\epsilon_R \ll 1$. With eqs.~\eqref{eq:widthSM} and \eqref{eq:widthDM}, it follows that
\begin{align}
\label{eq:relic_scaling_w_kappa}
\Omega h^2 \sim \frac{m_A'^2}{\kappa^2}
\end{align}
as long as the invisible $A'$ width dominates over the visible width. If the visible width is much larger than the invisible width, on the other hand, the relic density becomes independent of $\kappa$. The point where the two partial widths are equal, which marks the transition between the two regimes, is indicated by the dotted line in figure~\ref{fig:exp_excl_fixedeps}. This line moves upwards to larger $\kappa$ as $\gamma_\mathrm{inv}$ increases.

For $\gamma_\mathrm{inv}=10^{-9}$ (centre row of figure~\ref{fig:exp_excl_fixedeps}) a larger part of the shown parameter space yields primarily invisible $A'$ decays and the scaling in eq.~\eqref{eq:relic_scaling_w_kappa} holds true over the entire mass range from $10$~MeV to $10$~GeV. At the same time, bounds on invisibly decaying dark photons from NA64 replace beam dump searches relying on visible decays as the dominant constraint on dark photons with small mass.

Finally, for $\gamma_\mathrm{inv} = 10^{-5}$ (bottom row of figure~\ref{fig:exp_excl_fixedeps}), which corresponds to values of $g_\chi$ between $0.01$ and $0.1$, invisible $A'$ decays dominate over visible decays in nearly the entire parameter space. The strongest constraint on dark photons with mass below $200$~MeV again comes from NA64. In addition, the BaBar single-photon search excludes values of $\kappa \gtrsim 0.001$ for all masses below approximately $8$~GeV. While these experimental searches constrain $\kappa$ from above, CMB constraints come in from below. The latter are stronger for small $\kappa$ because they put a bound on $R^2 \langle \sigma v \rangle_\mathrm{CMB}$, which according to eq.~\eqref{eq:relic_scaling_w_kappa} and \eqref{eq:sigmav_CMB} scales like
\begin{align}
R^2 \langle \sigma v \rangle_\mathrm{CMB} \sim \frac{1}{\kappa^4} \, g_\chi^2 \kappa^2 = \frac{g_\chi^2}{\kappa^2}
\end{align}
when DM annihilations during freeze-out are resonantly enhanced while annihilations during or after recombination are not. As a consequence, for large $\gamma_\mathrm{inv}$ CMB constraints rule out large parts of the sub-GeV parameter space that yields the entire observed relic density. As we decrease $\epsilon_R$, CMB constraints are first weakened as annihilations during freeze-out become more strongly enhanced and the relic density at fixed $\kappa$ and $g_\chi$ decreases. However, for very small $\epsilon_{R}$, annihilations during recombination become resonantly enhanced as well, resulting in the stronger CMB constraints shown for $\epsilon_{R}=0.001$ (see figure~\ref{fig:sigmav}).

\subsection{Parameter scans}

As discussed in section~\ref{sec:accelerator}, the constraints from accelerator experiments shown in figure~\ref{fig:exp_excl_fixedeps} do not depend on $\epsilon_R$. Rather than fixing $\epsilon_R$ in each panel, we can therefore consider a range of different values at once. From the theoretical point of view there are no strict upper or lower bound on $\epsilon_R$, even though tiny values of $\epsilon_R$ require substantial fine-tuning and are possibly less attractive. Cosmological constraints, on the other hand, disfavour both very small values of $\epsilon_R$ (which would lead to unacceptably large annihilation rates during BBN) and very large values of $\epsilon_R$ (which would lead to unacceptably large annihilation rates during recombination). In the following, we therefore focus on the range $\epsilon_R \in [0.001, 1]$. We will discuss the impact of this restriction below and provide additional plots for $\epsilon_R \in [0.01, 1]$ in appendix~\ref{sec:appendix_eps}.

For each value of $\kappa$, $m_{A'}$ and $\gamma_\text{inv}$ we scan over $\epsilon_R$ and determine all values that satisfy the following requirements:
\begin{itemize}
 \item The predicted relic density satisfies $\Omega h^2 \leq 0.12$.
 \item The annihilation rate during recombination satisfies the bound discussed in section~\ref{sec:CMB}.
 \item The inferred value of $g_\chi$ satisfies the perturbativity requirement $g_\chi < \sqrt{4\pi}$.
 \item The DM self-interaction cross section satisfies the bound discussed in section~\ref{sec:SIDM}.
\end{itemize}
In practice, we find that points that satisfy the first three requirements also satisfy the last one, i.e.\ parameter points that predict large effects from self-interactions are already excluded by the other constraints. Note that we do not yet include the constraints from direct detection experiments, which will be discussed in detail below.

\begin{figure}[t]
	\centering
	\includegraphics[height=6.5cm,clip,trim=15 5 90 5]{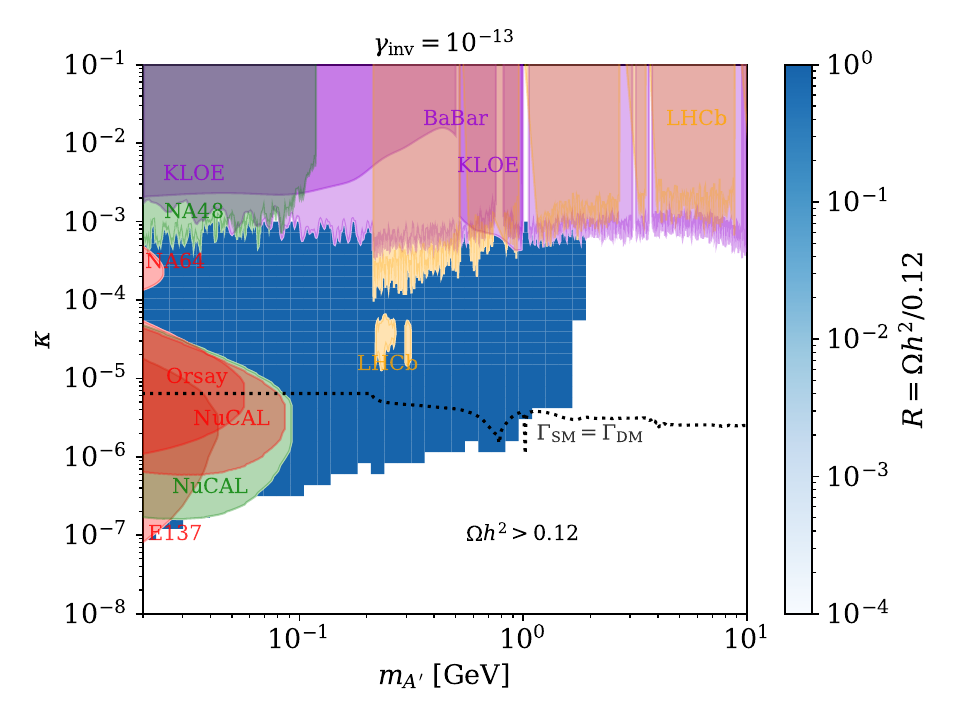}
	\includegraphics[height=6.5cm,clip,trim=15 5 15 5]{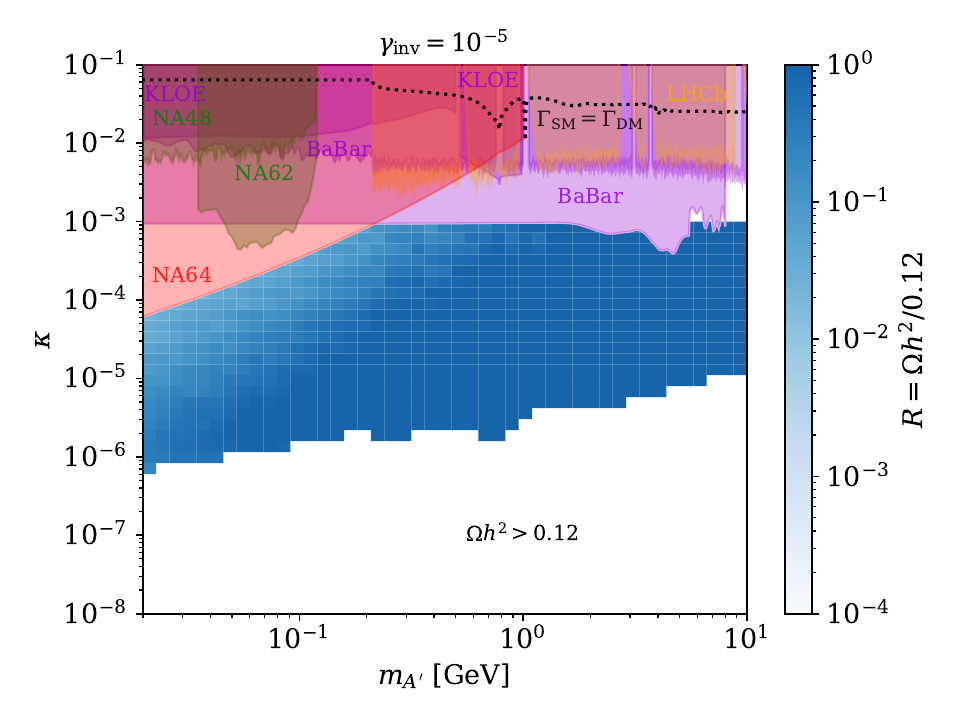}
	\caption{Same as figure~\ref{fig:exp_excl_fixedeps} but without holding $\epsilon_R$ fixed. Instead, we profile over $\epsilon_R$ by showing the largest value of $\Omega h^2$ that is allowed by all constraints.\label{fig:acc_excl}}
\end{figure}

The result of this procedure is shown in figure~\ref{fig:acc_excl} with the shaded regions indicating the parameter combinations for which viable values of $\epsilon_R$ can be found. The colour of the shading indicates the \emph{largest} relic abundance predicted by any of the allowed values of $\epsilon_R$. For example, if a point is shaded in dark blue, there exists (at least) one value of $\epsilon_R$ consistent with all constraints for which the DM particle would be all of DM. For points that are shown in white, all considered values of $\epsilon_R$ are excluded by at least one of the imposed constraints.

For the first case ($\gamma_\text{inv} = 10^{-13}$) the result of this procedure is fairly obvious: The allowed parameter regions in the $m_{A'}$--$\kappa$ plane correspond to those found for $\epsilon_R = 0.001$ in figure~\ref{fig:exp_excl_fixedeps}, and for all allowed points it is possible to find a value of $\epsilon_R$ such that the DM candidate constitutes all of DM. The boundary between the blue and the white regions is determined by the lower boundary on $\epsilon_R$ that we impose. If we were to include even smaller values of $\epsilon_R$ in our scan, the blue region would extend even further to the right and to the bottom. However, in this case it would no longer be appropriate to neglect BBN constraints on photodisintegration. We leave a more detailed analysis of this highly tuned parameter region to future work.

The second case ($\gamma_\text{inv} = 10^{-5}$) turns out to be much more interesting. As already seen in figure~\ref{fig:exp_excl_fixedeps}, for the case of large invisible width, CMB constraints are highly relevant for small $\kappa$ and determine the lower bound of the allowed parameter space. Since these constraints become stronger for smaller $\epsilon_R$, extending the range of $\epsilon_R$ below $0.001$ would not increase the shaded region. The second important observation is that it is no longer possible to saturate the relic density bound everywhere. For large $\kappa$ and small $m_{A'}$ all values of $\epsilon_R$ considered predict a subdominant DM component.

Figure~\ref{fig:acc_excl} hence reveals a strong complementarity between the different constraints: While the CMB constraint becomes stronger for smaller values of $\kappa$ (corresponding to larger abundances of the DM particle), accelerator constraints become stronger for larger values of $\kappa$. The fact that the DM particle constitutes only a sub-dominant DM component in these parameter regions is not an obstacle for these experiments, because they rely on the production (rather than the detection) of DM particles.

\bigskip

We now want to apply the procedure used to scan over $\epsilon_R$ also to the other model parameters. In this case, it is however no longer possible to show experimental constraints in the $m_{A'}$--$\kappa$ parameter plane, which can only be calculated for fixed values of $\gamma_\text{inv}$. Instead, we can show the viable parameter space of our model in the parameter plane relevant for direct detection experiments, i.e.\ the DM-electron scattering cross section rescaled by the fractional relic abundance, $R \, \sigma_e$, versus the DM mass, $m_\chi$. To find viable parameter points, we scan over the following parameter ranges:
\begin{itemize}
 \item $m_\chi \in [10\,\mathrm{MeV},5\,\mathrm{GeV}$], where the lower bound is dictated by BBN, while the upper bound reflects the focus of the present work on DM models below the threshold of conventional direct detection experiments;
 \item $\gamma_\text{inv} \in [10^{-23}, 0.1]$, where the upper bound is dictated by the perturbativity constraint and the lower bound is chosen such that all interesting values of $g_\chi$ are captured;
 \item $\kappa \in [10^{-10}, 0.001]$ where the upper bound reflects the bounds from accelerator experiments and the lower bound is chosen such that all interesting values of $\kappa$ are captured;
 \item $\epsilon_R \in [0.001, 1]$ for the reasons discussed above.
\end{itemize}

We note in passing that for the smallest couplings considered it is doubtful that the DM particles can maintain kinetic equilibrium during thermal freeze-out~\cite{Binder:2017rgn,Bringmann:2020mgx} and more refined methods for calculating the DM relic density may be necessary. However, we find that such small values of $\kappa$ and/or $g_\chi$ are of little practical interest, as they typically predict unacceptably large relic densities. This conclusion might however change when considering more finely tuned values of $\epsilon_R$. For the largest couplings that we consider, sizeable theory uncertainties arise from the fact that we neglect higher-order corrections and that we use a fixed instead of a running width for the dark photon propagator. These couplings, however, are typically either excluded by the CMB constraints or predict tiny relic densities and therefore do not notably affect the plots that we show.

\begin{figure}
\centering
\includegraphics[trim= 15 15 15 10, clip, width=0.65\textwidth ]{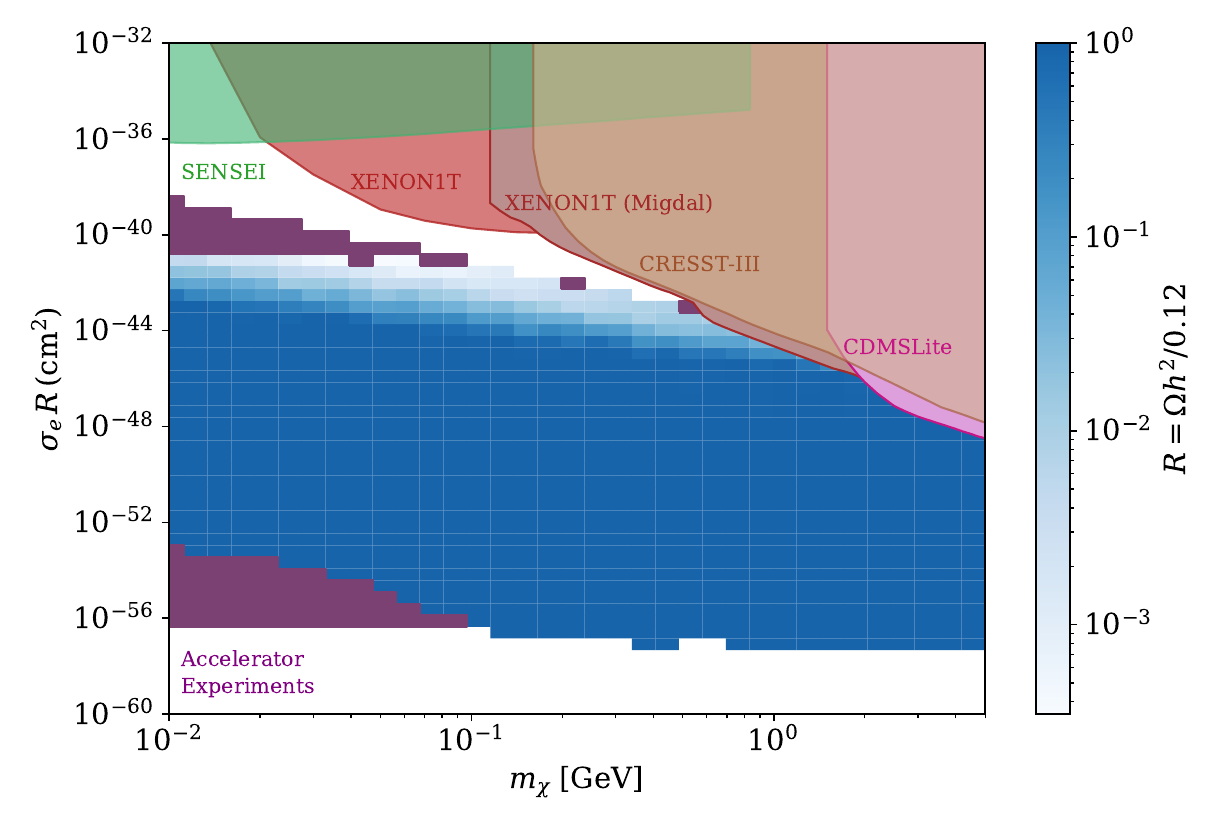}
\caption{Viable parameter space in the parameter plane relevant for direct detection experiments, i.e.\ the DM mass $m_\chi$ versus the DM-electron scattering cross section $\sigma_e$ multiplied with the fractional abundance of DM $R = \Omega h^2/0.12$. We also show existing bounds from direct detection experiments searching for DM-electron or DM-nucleus scattering. Purple shading indicates the parameter regions that are fully excluded by accelerator experiments.\label{fig:DD_bounds}}
\end{figure}

In figure~\ref{fig:DD_bounds} we show the viable parameter space in our model in the direct detection plane. As in figure~\ref{fig:acc_excl}, the shading represents the largest relic abundance predicted by any viable combination of parameters. In the parameter regions without shading, no viable parameter points survive after applying the four selection requirements listed above. Purple shading indicates those parameter regions for which no viable parameter points survive after applying the accelerator constraints.

Following our general discussions in section~\ref{sec:model}, the observed relic density requires smaller (larger) couplings when $\epsilon_R \ll 1$ ($\epsilon_R\to1$). Since $\sigma_e \propto \kappa^2 g_\chi^2$, moving from top to bottom in figure~\ref{fig:DD_bounds} is therefore equivalent to decreasing $\epsilon_R$ or moving closer to resonance in the model.

For small couplings, we found in the previous section that the strongest accelerator constraints (coming from fixed-target experiments) exclude dark photon masses up to a few hundred MeV. These constraints translate into the purple shaded region seen in the bottom left of the plot. 
For large couplings, there are two sets of constraints to keep in mind. First, the CMB constraints imply that we cannot saturate the relic abundance for large $\epsilon_R$, corresponding to small resonant enhancement. This results in a decrease in $R$ as we move upwards in the plot, as indicated by the lighter shading. Second, fixed-target experiments and $e^+e^-$ colliders provide additional constraints at large couplings, which reduce the viable parameter space.

We emphasize that accelerator bounds also exclude points along the hidden dimensions for intermediate couplings and cross sections, but it is always possible to find other viable parameter points mapping on to the same $R\sigma_e$ and hence these exclusions are not visible.   

As becomes immediately clear from figure~\ref{fig:DD_bounds}, direct detection constraints only play a minor role in constraining the parameter space of our model. Direct detection experiments sensitive to DM-electron scattering are not presently sensitive to the allowed regions of parameter space, while low-threshold experiments searching for nuclear recoils only exclude a small parameter region at large DM masses. Clearly, there is a substantial viable parameter space in the model that we consider. Even though we have only considered moderate resonant enhancement ($\epsilon_R \geq 0.001$), the model predictions cover more than ten orders of magnitude in terms of the rescaled DM-electron scattering cross section. In fact, this result is largely independent of the choice of the lower bound of $\epsilon_R$. As shown in appendix~\ref{sec:appendix_eps}, the allowed parameter space only shrinks insignificantly when requiring $\epsilon_R \geq 0.01$ instead.

\begin{figure}[t]
	\centering
	\includegraphics[height=6.5cm,clip,trim=15 5 90 5]{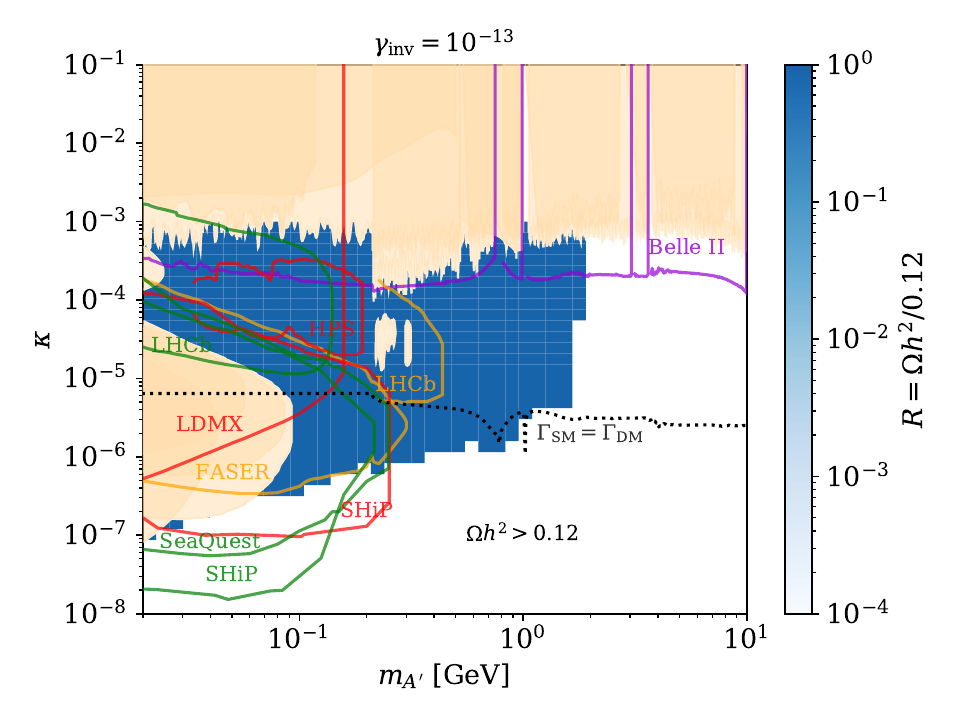}
	\includegraphics[height=6.5cm,clip,trim=15 5 15 5]{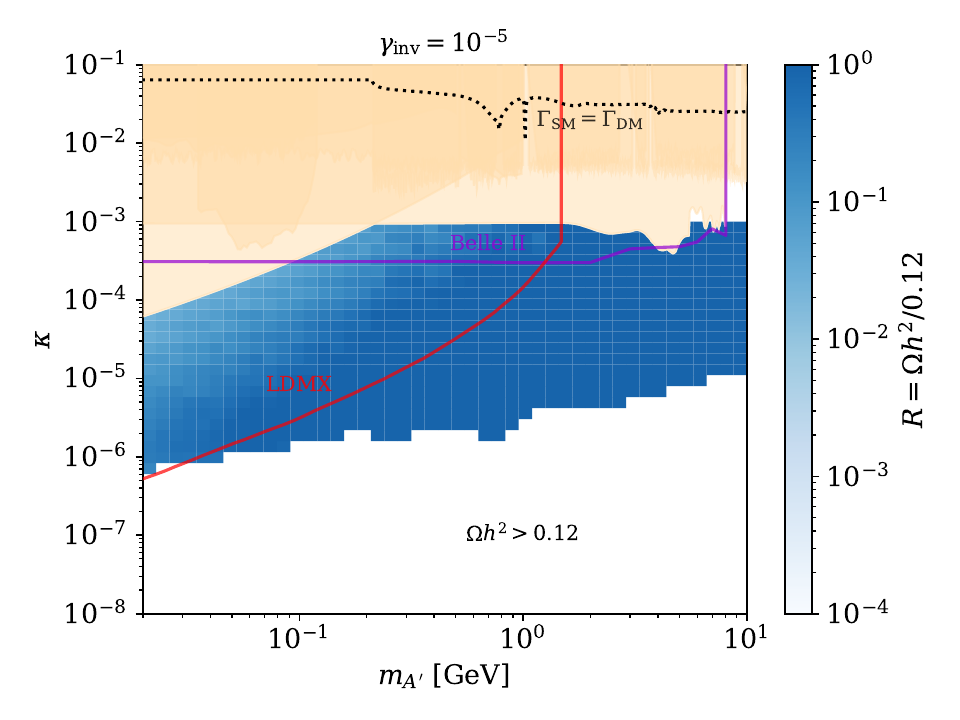}
	\caption{Same as figure~\ref{fig:acc_excl} but with existing constraints shaded in orange and projected sensitivities indicated by solid lines. The line colours depend on the dark photon production mechanism as in figures~\ref{fig:exp_excl_fixedeps} and \ref{fig:acc_excl}. \label{fig:acc_proj}}
\end{figure}

\begin{figure}
\centering
\includegraphics[trim= 15 15 15 10, clip,width=0.65\textwidth]{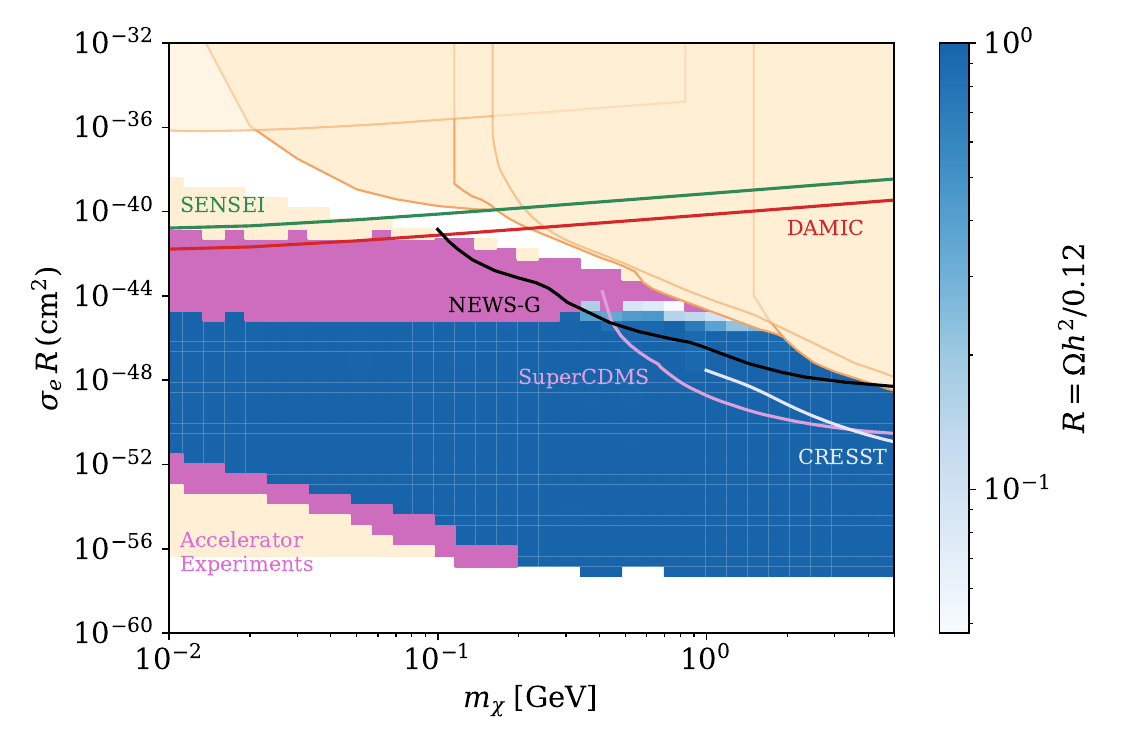}
\caption{Same as figure~\ref{fig:DD_bounds} but with existing constraints shaded in orange and projected sensitivities indicated by solid lines. The magenta shaded region can be fully explored by future accelerator experiments. \label{fig:DD_proj}}
\end{figure}

To understand how the next generation of experiments will have an impact on these conclusions, we show in figures~\ref{fig:acc_proj} and \ref{fig:DD_proj} various projected sensitivities.\footnote{We note that the different projections that we show range from future updates of already running experiments over experiments that are under construction to some that have not yet secured funding.} For the case of a small invisible width (left panel of figure~\ref{fig:acc_proj}) a whole range of experiments will thoroughly explore the parameter space for visibly decaying sub-GeV dark photons, with the strongest projected bounds coming from Belle II~\cite{Kou:2018nap}, FASER~\cite{Ariga:2018uku}, HPS~\cite{Baltzell:2016eee}, LHCb~\cite{Ilten:2015hya, Ilten:2016tkc}, SeaQuest~\cite{Gardner:2015wea} and SHiP~\cite{Alekhin:2015byh}. For the case of dominant invisible decays (right panel of figure~\ref{fig:acc_proj}) improved single-photon and missing energy searches again promise substantial progress in the search for dark photons, most notably at Belle II and LDMX~\cite{Akesson:2018vlm, Beacham:2019nyx}.

The parameter regions that can be fully explored by future accelerator experiments are shaded in magenta in figure~\ref{fig:DD_proj}. We find that these developments are highly complementary to the expected advances in direct detection experiments. While accelerator experiments are particularly sensitive to small dark photon (and hence DM) masses, direct detection experiments searching for low-energy nuclear recoils are expected to make substantial progress for DM masses in the GeV range. Interestingly, these experiments will first probe parameter regions in which the DM particle is predicted to only be a sub-component of the total DM density.

Still, there remain large regions of parameter space that are extremely challenging to probe experimentally. The sensitivity of direct detection experiments is limited by the background from coherent scattering of solar neutrinos, and fixed-target experiments struggle to probe dark photon masses above the GeV scale. As shown in appendix~\ref{sec:appendix_eps}, these conclusions are almost completely insensitive to the precise lower bound on $\epsilon_R$ imposed in the analysis. We conclude that even in very simple models without excessive fine-tuning the freeze-out mechanism remains viable and new experimental strategies will be needed to fully explore the parameter space.

\section{Conclusions}
\label{sec:conclusions}

In this work we have conducted a global analysis of a sub-GeV Dirac DM model with a dark photon mediator. To evade the strong cosmological constraints on light thermal particles, we consider a scenario where the DM candidate is approximately half as massive as the dark photon. In this case, the annihilations, $\mathrm{DM}\,\mathrm{DM}\to A' \to \mathrm{SM}\,\mathrm{SM}$, are resonantly enhanced at large temperatures, as a result of which the observed DM relic abundance can be obtained for rather tiny couplings. 
These small couplings ensure that DM annihilations are sufficiently suppressed during BBN and recombination to satisfy the respective cosmological bounds. As a second way of suppressing cosmological constraints we point out the possibility that the sub-GeV DM particle constitutes only a DM subcomponent. This expands the viable parameter space significantly while not adversely affecting the testability of this model at current and future terrestrial experiments. We also calculate the constraints on the DM self-interaction and show that these are generally weaker than those coming from the CMB. 

We evaluate the cosmological and terrestrial bounds using state-of-the-art numerical tools. In particular, we use injection spectra calculated with \textsc{HERWIG4DM} to calculate accurate CMB constraints for GeV-scale DM particles annihilating into quarks and leptons and extend the recasting tool \textsc{Darkcast} to the case of dark photons with arbitrary invisible branching ratios. Using these results we demonstrate the complementarity between cosmological constraints (which exclude parameter points with small resonant enhancement), accelerator searches for dark photons (which can probe relatively small couplings and hence strong resonant enhancement) and direct detection constraints (which turn out to be particularly sensitive to GeV-scale DM and DM sub-components).

Our key results are summarised in figures~\ref{fig:acc_proj} and~\ref{fig:DD_proj}, which show the viable parameter space in the plane relevant for accelerator experiments and direct detection experiments, respectively. To obtain these figures, we have scanned over the full parameter space of the model and identified the points that predict the largest DM relic abundance consistent with all constraints. We note that this approach is possible only because all constraints are extremely fast to evaluate. In particular, we have made use of several different analytic approximations for the thermally averaged DM annihilation cross section in order to substantially speed up the calculation of the DM relic abundance.

In summary, although the model is constrained by many different experiments and observations, we find large allowed regions of parameter space. Our results demonstrate that even simple models of thermal DM remain viable with only moderate amounts of tuning. Moreover, we highlight the great potential of terrestrial experiments to search for DM sub-components, which readily evade cosmological constraints. While near-future experiments will probe deeper into the allowed parameter space, innovative and ambitious strategies will be needed to thoroughly explore the freeze-out paradigm.

It will be interesting to extend the analysis presented here to even stronger resonant enhancement and hence even smaller couplings. In this case a number of new considerations will become relevant. First of all, the resonant enhancement may be non-negligible during BBN, such that constraints from photodisintegration need to be taken into account. Furthermore, for very small couplings supernova constraints become important, which need to be re-evaluated for the case that the dark photon can decay into DM particles (which may then get trapped inside the supernova). Finally, the relic density calculation itself may be modified, because the dark sector can no longer maintain kinetic equilibrium during freeze-out. The technical developments needed to address these questions offer promising directions for future work.

\acknowledgments
We thank Michael Kr\"{a}mer, Tilman Plehn, Peter Reimitz, Kai Schmidt-Hoberg and Patrick St\"{o}cker for discussions and Peter Reimitz for providing energy spectra from \textsc{HERWIG4DM}. This  work  is  funded  by  the  Deutsche Forschungsgemeinschaft (DFG) through the Collaborative Research Center TRR 257 ``Particle Physics Phenomenology after the Higgs Discovery'' under Grant  396021762 - TRR 257 and the Emmy Noether Grant No.\ KA 4662/1-1.

\begin{appendix}
	\section{Recasting of accelerator constraints}
	\label{sec:appendix_recasting}
	In recasting limits for our model we follow the procedure in ref.~\cite{Ilten:2018crw} implemented in the public companion code \textsc{Darkcast}, which we have modified to correctly take into account the invisible width of the dark photon (eq.~\eqref{eq:widthDM}) in the calculation of its branching ratios and lifetime.
	The number of signal events $N$ at a fixed-target experiment or collider predicted by a given dark photon model is proportional to the product of the production cross section $\sigma_{A'}$, the branching ratio $\mathrm{BR}_{A'\to\mathcal{F}}$ of $A'$ into the relevant final state and the experimental efficiency $\epsilon$:
	\begin{align}
	N \propto \sigma_{A'} \; \mathrm{BR}_{A'\to\mathcal{F}} \; \epsilon \; .
	\end{align}
	
	Assuming full efficiency everywhere in the fiducial volume, the overall efficiency $\epsilon$ is simply given by the probability that the dark photon decays within the fiducial volume, i.e.\ after travelling a distance between $L_\mathrm{sh}$ and $L_\mathrm{sh}+L_\mathrm{dec}$ in the laboratory frame. Here $L_\mathrm{sh}$ denotes the length of the shielding and $L_\mathrm{dec}$ the length of the decay volume for a given experiment. For the decay to happen in this distance interval, the particle has to decay between the times $\tilde{t}_0$ and $\tilde{t}_1$ in its rest frame, where $\tilde{t}_0 = L_\mathrm{sh}/(c\beta\gamma)$ and $\tilde{t}_1 = (L_\mathrm{sh}+L_\mathrm{dec})/(c\beta\gamma)$. The efficiency for a dark photon with lifetime $\tau$ is then given by
	\begin{align}
	\label{eq:efficiency}
	\epsilon(\tau) = e^{-\tilde{t}_0/\tau} - e^{-\tilde{t}_1/\tau} \; .
	\end{align}
	Assuming approximately mono-energetic production of dark photons with a given mass $m_{A'}$ in a given experiment, the times $\tilde{t}_0$ and $\tilde{t}_1$ can be determined for each $m_{A'}$ using the following relations. First, their ratio is always equal to
	\begin{align}
	\label{eq:time_ratio}
	\frac{\tilde{t}_1}{\tilde{t}_0} = \frac{L_\mathrm{sh}+L_\mathrm{dec}}{L_\mathrm{sh}} \; .
	\end{align}
	Second, the lowest and highest mixing parameter $\varepsilon$ excluded by the experiment for some $m_{A'}$ both predict the same number of signal events, which is equal to the statistical upper limit $N_\mathrm{excl}$. Hence it follows that
	\begin{align}
	\label{eq:upper_lower_equality}
	\varepsilon^2_\mathrm{max} \, \epsilon(\tau_{A'_0}(\varepsilon^2_\mathrm{max})) = \varepsilon^2_\mathrm{min} \, \epsilon(\tau_{A'_0}(\varepsilon^2_\mathrm{min})) \; ,
	\end{align}
	where $A'_0$ denotes the dark photon in the model considered by the experiment.
	The values of $\tilde{t}_0$ and $\tilde{t}_1$ as a function of $m_{A'}$ are then determined by inserting the efficiency from  eq.~\eqref{eq:efficiency} in eq.~\eqref{eq:upper_lower_equality} and combining with eq.~\eqref{eq:time_ratio}.
	
	When $\tilde{t}_0$ and $\tilde{t}_1$ are known, the boundaries of the excluded region can then be recast by determining the $\kappa_{\mathrm{min}/\mathrm{max}}(m_{A'})$ for which our model predicts the same number of events as the original model does for $\varepsilon_{\mathrm{min}/\mathrm{max}}(m_{A'})$, i.e.\ by solving
	\begin{align}
	\left[\sigma_{A'} \; \mathrm{BR}_{A'\to\mathcal{F}} \; \epsilon(\tau_{A'})\right]_{\kappa_{\mathrm{min}/\mathrm{max}}} = \left[\sigma_{A'_0} \; \mathrm{BR}_{A'_0\to\mathcal{F}} \; \epsilon(\tau_{A'_0})\right]_{\varepsilon_{\mathrm{min}/\mathrm{max}}}
	\end{align}
	for $\kappa_{\mathrm{min}/\mathrm{max}}$ for each mass $m_{A'}$.

	Note that estimating the efficiency in this way is not necessary for recasting the limits of the LHCb search for displaced decays in ref.~\cite{Aaij:2017rft}, since the ratio $r_\mathrm{ul}^\mathrm{ex}$ of the experimental upper bound on the number of dark photon decays to the expected number of dark photon decays is provided as a function of $m_{A'}$ and $\varepsilon$. With this information, \textsc{Darkcast} can simply exclude a parameter point in our dark photon model when
	\begin{align}
	\left[ r_\mathrm{ul}^\mathrm{ex}(m_{A'}, \varepsilon) \, \frac{\sigma_{A'_0} \; \mathrm{BR}_{A'_0\to\mathcal{F}}}{\sigma_{A'} \; \mathrm{BR}_{A'\to\mathcal{F}}} \right]_{\tau_{A'}=\tau_{A'}}  < 1\; .
	\end{align}

\begin{figure}[t]
	\centering
	\includegraphics[height=6.5cm,clip,trim=15 5 90 5]{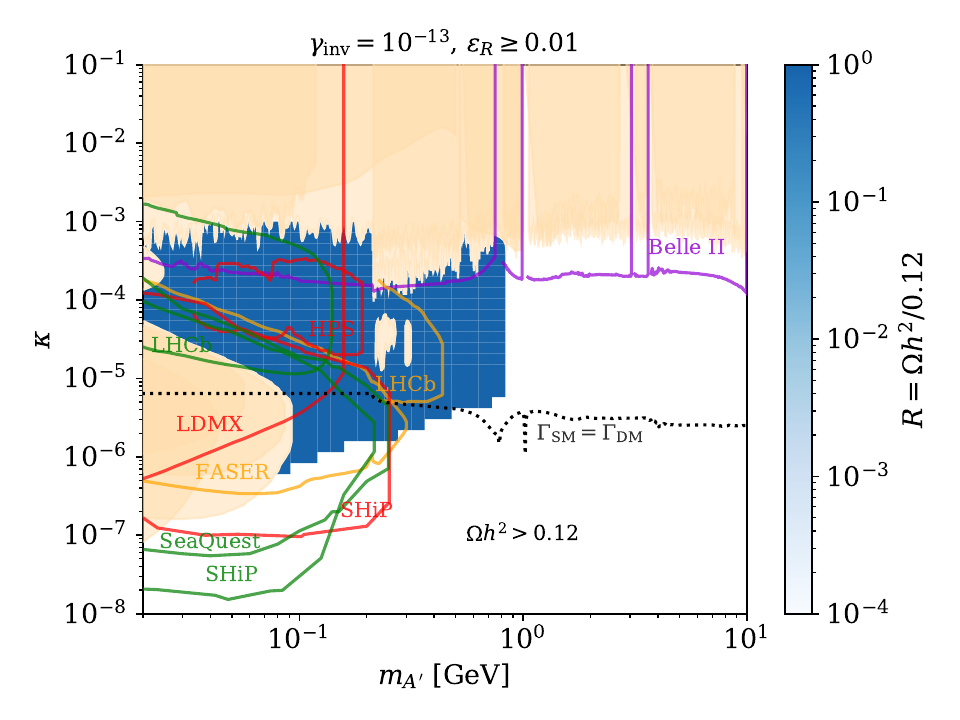}
	\includegraphics[height=6.5cm,clip,trim=15 5 15 5]{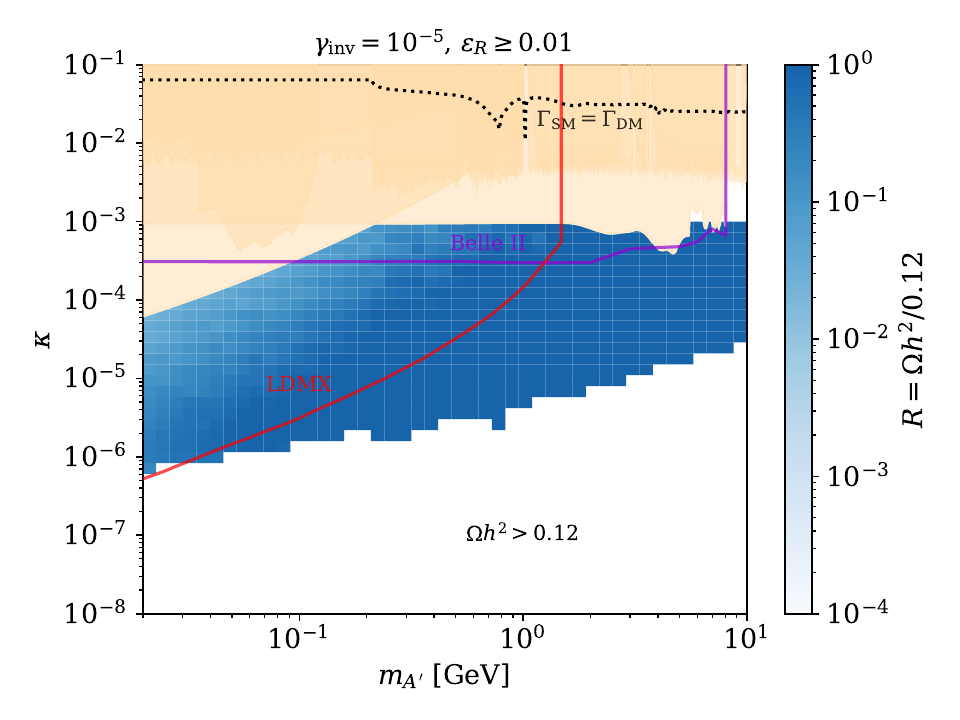}
	\caption{Same as figure~\ref{fig:acc_proj} but with the restriction $\epsilon_R \geq 0.01$. \label{fig:acc_proj_large_eps}}
\end{figure}

\section{Parameter space for $\epsilon_R \geq 0.01$}
\label{sec:appendix_eps}

\begin{figure}
\centering
\includegraphics[trim= 15 15 15 10, clip, width=0.65\textwidth ]{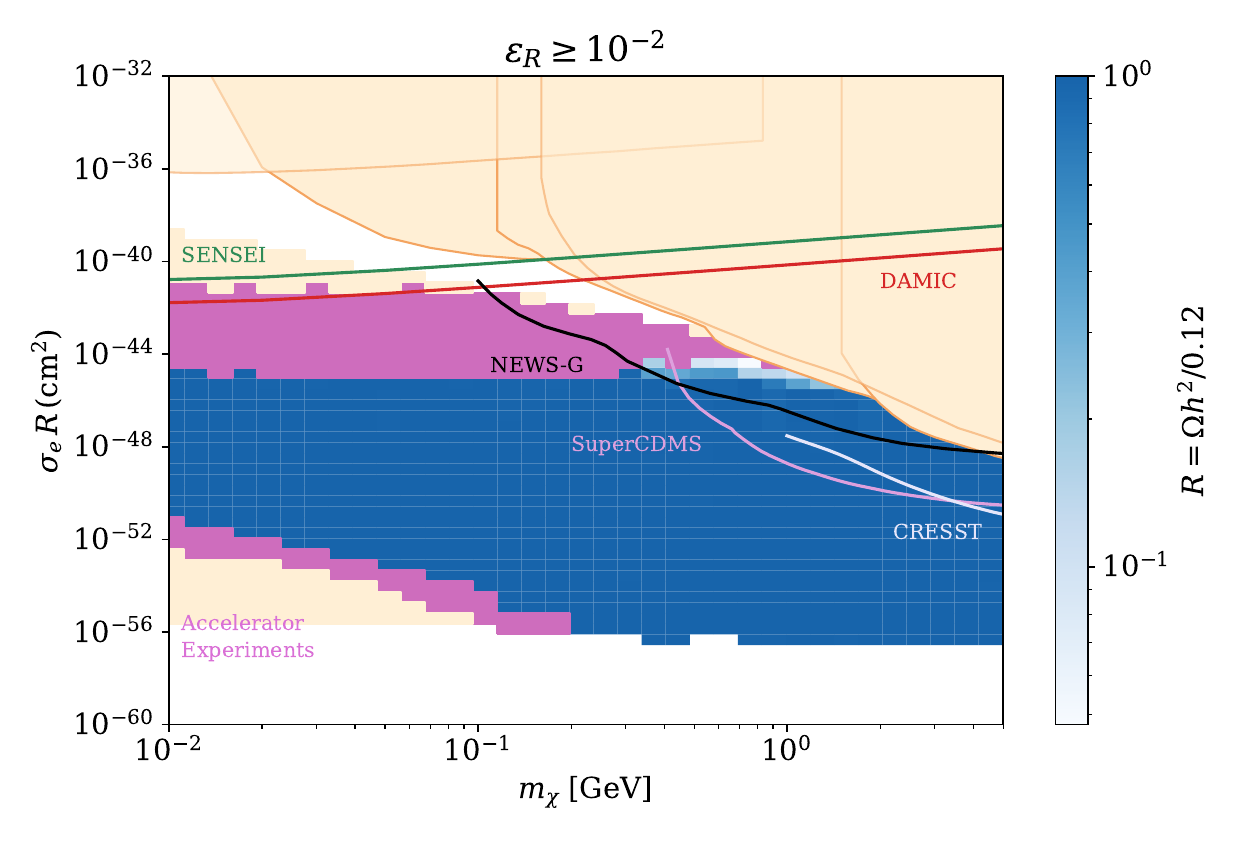}
\caption{\label{fig:DD_proj_large_eps} Same as figure~\ref{fig:DD_proj} but with the restriction $\epsilon_R \geq 0.01$.}
\end{figure}	

In figures~\ref{fig:acc_proj_large_eps} and~\ref{fig:DD_proj_large_eps}, we show the viable parameter space for $\epsilon_R \geq 0.01$ with the same conventions as discussed in section~\ref{sec:results}. We find that restricting the range of $\epsilon_R$ to larger values has almost no impact on our results. While the lower boundary of the allowed parameter space shifts slightly, the upper regions, which are in reach of future experiments, remain completely unaffected. This implies that no excessive fine-tuning is necessary for the model to remain viable and that its detection prospects are insensitive to the smallest values of $\epsilon_R$.

\end{appendix}

\bibliographystyle{JHEP_improved}
\bibliography{biblio.bib}

\end{document}